\documentclass[pre,aps,reprint,superscriptaddress,titlepage,floatfix,showkeys]{revtex4-1}
\usepackage{amsmath}
\usepackage{bm}             
\usepackage[utf8]{inputenc}
\usepackage{latexsym, amsfonts, amssymb, wasysym}
\usepackage[dvipsnames, table]{xcolor}
\usepackage{graphicx}
\usepackage{nicefrac}
\usepackage{amsfonts}
\usepackage[version=3]{mhchem}
\usepackage{subfigure}
\usepackage{multirow}
\usepackage{tabularx}
\usepackage{array}
\usepackage{units}
\usepackage{braket}
\usepackage{bm,times}
\usepackage{booktabs}
\usepackage{enumitem}
\usepackage{mathtools}
\usepackage{epstopdf}
\usepackage{threeparttable}
\usepackage{tensor} 

\usepackage[colorlinks = true, linkcolor = blue, urlcolor  = blue, citecolor = blue, anchorcolor = blue]{hyperref}

\def\be{\begin{equation}}
\def\ee{\end{equation}}
\def\bea{\begin{eqnarray}}
\def\eea{\end{eqnarray}}

\def\degree{$^\circ$}

\def\asi{{\it a}-Si}
\def\age{{\it a}-Ge}

\def\vsio2{{\it v}-SiO$_2$} 
\def\AAI{\AA$^{-1}$}
\def\lena{\langle R_2^3\rangle}
\def\lenb{\langle R_4^3\rangle}

\definecolor{lightgray}{gray}{0.85}
\newcolumntype{a}{>{\columncolor{lightgray}}c}

\definecolor{lightgray}{gray}{0.9}
\definecolor{dgray}{gray}{0.6}
\definecolor{lblue}{rgb}{0.67, 0.9, 0.93}
\definecolor{salmon}{rgb}{1, 0.6, 0.6}
\definecolor{green1}{rgb}{0.47, 0.87, 0.47}
\definecolor{pyellow}{rgb}{0.99, 0.99, 0.59}

\begin{document}

\title{On the origin and structure of the first sharp diffraction peak of amorphous silicon}

\author{Devilal Dahal}
\affiliation{Department of Physics and Astronomy, The University of
Southern Mississippi, Hattiesburg, Mississippi 39406, USA}

\author{Hiroka Warren} 
\affiliation{Department of Physics and Astronomy, The University of
Southern Mississippi, Hattiesburg, Mississippi 39406, USA}

\author{Parthapratim Biswas}
\email[To whom correspondence should be addressed:\,]{partha.biswas@usm.edu} 
\affiliation{Department of Physics and Astronomy, The University of
Southern Mississippi, Hattiesburg, Mississippi 39406, USA}

\begin{abstract}
The structure of the first sharp diffraction peak (FSDP) of 
amorphous silicon ({\it a}-Si) near 2 {\AA}$^{-1}$ is 
addressed with particular emphasis on the position, intensity, 
and width of the diffraction curve. By studying a number of 
continuous random network (CRN) models of {\asi}, it is shown 
that the position and the intensity of the FSDP are primarily 
determined by radial atomic correlations in the amorphous 
network on the length scale of 15 {\AA}. A shell-by-shell 
analysis of the contribution from different radial shells 
reveals that key contributions to the FSDP originate from 
the second and fourth radial shells in the network, which 
are accompanied by a background contribution from the first 
shell and small residual corrections from the distant radial shells. 
The results from numerical calculations are complemented by 
a phenomenological discussion of the relationship between 
the peaks in the structure factor in the wavevector space 
and the reduced pair-correlation function in the real space. 
An approximate functional relation between the position of 
the FSDP and the average radial distance of Si atoms in 
the second radial shell in the network is derived, which 
is corroborated by numerical 
calculations. 
\end{abstract} 

\keywords{Amorphous silicon, Pair-correlation function, 
Static structure factor, First sharp diffraction peak}
\maketitle   

\section{Introduction}
\label{S:1}
Professor David Drabold has contributed significantly in the field 
of amorphous materials. It is therefore an opportune moment to contribute 
to his Festschrift on a topic which is very close to his heart. 
The first sharp diffraction peak (FSDP) is a distinct feature of 
many noncrystalline solids, which are characterized by 
the presence of a peak in the low wavevector region (1--2 {\AA}$^{-1}$) 
of the structure factor of the solids. Although the origin of 
the FSDP in many multinary glasses is not yet fully understood from 
an atomistic point of view, it has been shown that the FSDP is 
primarily associated with the presence of the short-range and 
medium-range order, which entail voids, chemical ordering, large 
ring structures, local topology, and atomic correlations between 
constituent atoms in the amorphous environment of the 
solids~\cite{Salmon:2005,Angelo:2010,Elliott:1995,Uhlherr:1995,Crupi:2016}. 

The FSDP is ubiquitous in many disordered condensed-phase systems. 
Numerous experimental~\cite{Salmon:2005,Angelo:2010,Bychkov:2005,Uemura:1975,Susman:1988} 
and theoretical~\cite{Elliott:1995,Vashishta:1989,Price:1989} 
studies have reported the (near) universal presence of the FSDP 
in glasses and liquids/melts. In glasses, the origin of the FSDP 
can be largely attributed to the presence of layered 
structures~\cite{Busse:1981}, interstitial 
voids~\cite{Elliott:1995,Uhlherr:1995,Crupi:2016}, chemical 
disorder~\cite{Uhlherr:1995}, and large ring 
structures~\cite{Susman:1988} in the networks, which constitute 
a real-space description of atomic correlations on the nanometer 
length scale. Elliott~\cite{Elliott:1995} has shown that 
the FSDP in binary glasses can be interpreted as 
a prepeak in the concentration-concentration structure factor, 
which is caused by the presence of the chemical ordering 
between constituent atoms in the networks. Likewise, the 
interstitial voids have been found to play an important role 
in the formation of the FSDP in tetrahedral amorphous 
semiconductors~\cite{Uhlherr:1995}, e.g., {\asi}. On the other 
hand, Susman et al.~\cite{Susman:1988} have reported that in 
binary AX$_2$ glasses, the A--A and A--X correlations within 
the extended ring structures can give rise to the FSDP. Busse 
and Nagel~\cite{Busse:1981} have suggested that the existence 
of the FSDP in {\it{g}}-As$_{2}$Se$_{3}$ can be ascribed to 
the inter-layer atomic correlations in the glassy network. 
Experimental studies on GeSe$_{3}$ and GeSe$_{5}$ glasses 
by Armand et al.~\cite{Armand:1994} have indicated that the 
Ge--Ge atomic correlation on the length scale of 6-7 {\AA} is 
the primary cause of the FSDP, which is supported by 
molecular-dynamics studies by Vashishta et al.~\cite{Vashishta:1989}.  

The behavior of the FSDP in covalent glasses often shows an 
anomalous dependence with respect to temperature\,\cite{Angelo:2010,Susman:1991}, 
pressure\,\cite{Susman:1991,Inamura:1997}, and 
composition~\cite{Bychkov:2005,Lee:1995}.  Following the 
well-known Debye-Waller~\cite{Debye:1913} behavior, one may assume that the
peaks in the structure factor should decrease with the increase of
the temperature of the system. However, the first (sharp) diffraction
peak of many glassy systems has been found to remain either invariant
or become more intense and narrower at high temperature~\cite{Angelo:2010,Inamura:2007}.  A notable exception is vitreous silica ({\vsio2}), which
does not follow the behavior stated above. The intensity of the FSDP
of {\vsio2} has been observed to decrease with increasing temperature,
due to the thermally induced motion of the atoms and the associated
diffused scattering~\cite{DWnote}, leading to the broadening of the
first peak~\cite{Susman:1991}. Likewise, the position and the width of
the FSDP have been observed to vary with the pressure or density of
the glasses~\cite{Inamura:2007,Inamura:1997}.
Neutron diffraction~\cite{Susman:1991_2} and
molecular-dynamics studies of densified {\vsio2}~\cite{Pilla:2003}
have indicated that the intensity and the width of the FSDP can change with the density of
the samples/models. These changes can be attributed to the
frustration induced by the reduction of Si--O--Si bond angles and
the changes in the Si--Si and O--O atomic correlations on the
length scale of 4--10 {\AA} when the system is densified.
A similar conclusion can be made for GeO$_{2}$ glass, when
the glass is densified~\cite{Sampath:2003}.
The  addition of extrinsic atoms in glassy networks has been also
found to affect the first sharp diffraction peak.  Lee and
Elliott~\cite{Lee:1995} have noted that the inclusion of extrinsic
atoms in {\vsio2} can change the chemical ordering of the
interstitial voids in the glassy network, which can alter
the shape/width of the FSDP.

While the great majority of earlier studies mostly examined 
the origin and the behavior of the FSDP in borate, chalcogenide, oxide, and 
silicate glasses~\cite{Crupi:2016,Lucovsky:2010,Gaskell:1996,Du:2005},  
there exist only a few studies~\cite{Elliott:1995,Uhlherr:1995,Uhlherr:1994} 
that address the structure of the FSDP in tetrahedral 
amorphous semiconductors, such as {\asi} and {\age}. 
Elliott and coworkers~\cite{Uhlherr:1995,Uhlherr:1994} have 
addressed the problem at length, but their studies are 
primarily focused on the origin of the extended-range 
oscillations (ERO) in {\asi}. The results from their 
studies, which are based on the (Fourier) inversion of 
experimental structure-factor data of Fortner and 
Lannin~\cite{Fortner:1989} and highly defective {\asi} 
models of Holender and Morgan~\cite{Holender:1991}, 
suggest that the ERO arise from the preferential 
propagation of second-neighbor correlations in the 
network, which in turn can significantly affect the 
intensity of the FSDP up to a radial length scale of 
20 {\AA}. 
However, no systematic analysis of the results with 
respect to the size of models is provided and, thus, 
in the absence of direct numerical evidence, it is 
not clear to what extent the intensity of the FSDP 
is truly affected by atomic correlations originating from 
radial distances beyond 15 {\AA}. 

The key purpose of this paper is to provide 
a systematic study of the structure of the FSDP, 
with an emphasis on the position, intensity, 
and width of the peak, with the size of the 
models. In addition, the origin of the FSDP in 
{\asi} is addressed by obtaining a quantitative 
estimate of the contribution of atomic 
pair correlations from different radial shells 
and their effect on the intensity and position 
of the FSDP. The relationship between the peaks 
in the structure factor and its real-space counterpart, 
the reduced pair-correlation function, is addressed, 
and an approximate functional relation between 
the position of the FSDP in {\asi} and the 
radial distance of the atoms in the second radial 
shell of the amorphous network is obtained. 
Throughout this paper, we shall use the term FSDP 
to refer to the first peak of the structure factor 
of {\asi} at $Q_0$ = 1.9--2 {\AA}$^{-1}$ in 
discussing our results.  Likewise, the term principal 
peak will be used to indicate the second peak at 
$Q$ = 3.6 {\AA}$^{-1}$. For amorphous silicon, 
this terminology has been used previously by 
others~\cite{Uhlherr:1995,Uhlherr:1994}, and it 
is consistent with the fact that the peak at 
$Q_0$ is indeed the first peak of $S(Q)$ and 
that it is reasonably sharp and strong with a 
value of the intensity $S(Q_0)$, which is about 
67\% of the intensity of the principal peak. 
A further justification of the use of the terminology 
will be evident later from our discussion of the 
results in section IIIA. 

The rest of the paper is planned as follows.  Section \ref{S:2} 
provides a brief description of the simulation method for producing 
atomistic models of {\asi} via the modified Wooten-Winer-Weaire 
(WWW)~\cite{Wooten:1985,Barkema:2000} method, 
the calculation of the radial pair-correlation function, 
and the structure factor for these models.  This is followed 
by results and discussion in section III, with an emphasis 
on the origin and the structure of the FSDP. The 
conclusions of this study are presented in section IV. 

\section{Models and Methods}
\label{S:2}
For the purpose of generating atomistic models of {\asi}, we 
have employed the well-known WWW method. The details of the 
method can be found in Refs.~\cite{Wooten:1985,Barkema:2000}. 
Here, we have used the modified version of the method, developed 
by Barkema and Mousseau (BM)~\cite{Barkema:2000}. In 
the modified WWW approach, one starts with a random 
configuration that consists of $N$ atoms in a cubic supercell 
of length $L$.  The volume of the supercell is chosen in 
such a way that the mass density of the model corresponds 
to about 2.28 g.cm$^{-3}$, as observed in {\asi} samples 
produced in laboratories~\cite{Xie:2013,Laaziri:1999}.  
Initially, following the BM ansatz, the nearest neighbors of each 
atom are so assigned that a tetravalent network is 
formed~\cite{BMnote}. This is achieved by choosing a 
suitable nearest-neighbor cutoff distance, up to 
3 {\AA}, between Si atoms. The resulting tetravalent 
network is then used as the starting point 
of the WWW bond-switching algorithm.  New configurations 
are generated by introducing a series of WWW bond switches, 
which largely preserve the tetravalent coordination of 
the network and the energy of the system is minimized 
using Monte Carlo (MC) simulations. The acceptance 
or rejection of a proposed MC move is determined via the 
Metropolis algorithm~\cite{Metropolis:1953} at a given 
temperature. Here, the energy difference between two 
configurations is calculated locally by using the Keating 
potential~\cite{Keating:1966}, which employs an 
atomic-index-based nearest-neighbor list of the 
tetravalent network during MC simulations. 
In addition, the total energy of the entire 
system is relaxed from time to time using the Stillinger-Weber 
potential~\cite{SW:1985}. Finally, the configurations obtained from 
the modified WWW method were relaxed using the first-principles 
density-functional code {\sc Siesta}~\cite{Soler:2002}. For 
the models with 216 atoms to 3000 atoms, a full self-consistent-field 
calculation, using the generalized-gradient 
approximation (GGA)~\cite{Perdew:1996} and a set of 
double-zeta basis functions, was carried out. 
The remaining models of size from 4096 
atoms to 6000 atoms were treated using the 
non-self-consistent Harris-functional approach~\cite{Harris:1985} 
with a single-zeta basis set in the local density 
approximation (LDA)~\cite{Perdew:1981}. To conduct 
configurational averaging of simulated data, we have 
generated 10 models for each size starting with 
different random configurations using independent 
runs.

Once the atomistic models are generated, the calculation of the 
structure factor proceeds by computing the reduced pair-correlation 
function. The latter is defined as $G(r)= 4\pi r n_0 [g(r)-1]$, 
where $g(r)$ and $n_0$ are the pair-correlation function 
and the average number density of a model, respectively.  Assuming 
that the distribution of the atoms in a disordered network is 
isotropic and homogeneous, the structure factor, $S(Q)$, can 
be written as,

\bea 
S(Q) &=& 1 + \frac{1}{Q} \int_0^{\infty}  G(r) \sin(Qr)\, dr \notag \\ 
     &\approx& 1 + \frac{1}{Q} \int_0^{R_c}  G(r) \sin(Qr)\, dr, 
\label{E1}
\eea
\noindent
where $R_c$ is the length of the half of the cubic simulation 
cell. The conventional periodic boundary conditions are 
used to minimize surface effects and to calculate the pair-correlation 
function in Eq.\,(\ref{E1}).

\section{Results and Discussion}
\begin{figure}[t!]
\centering
\includegraphics[width=.4\textwidth]{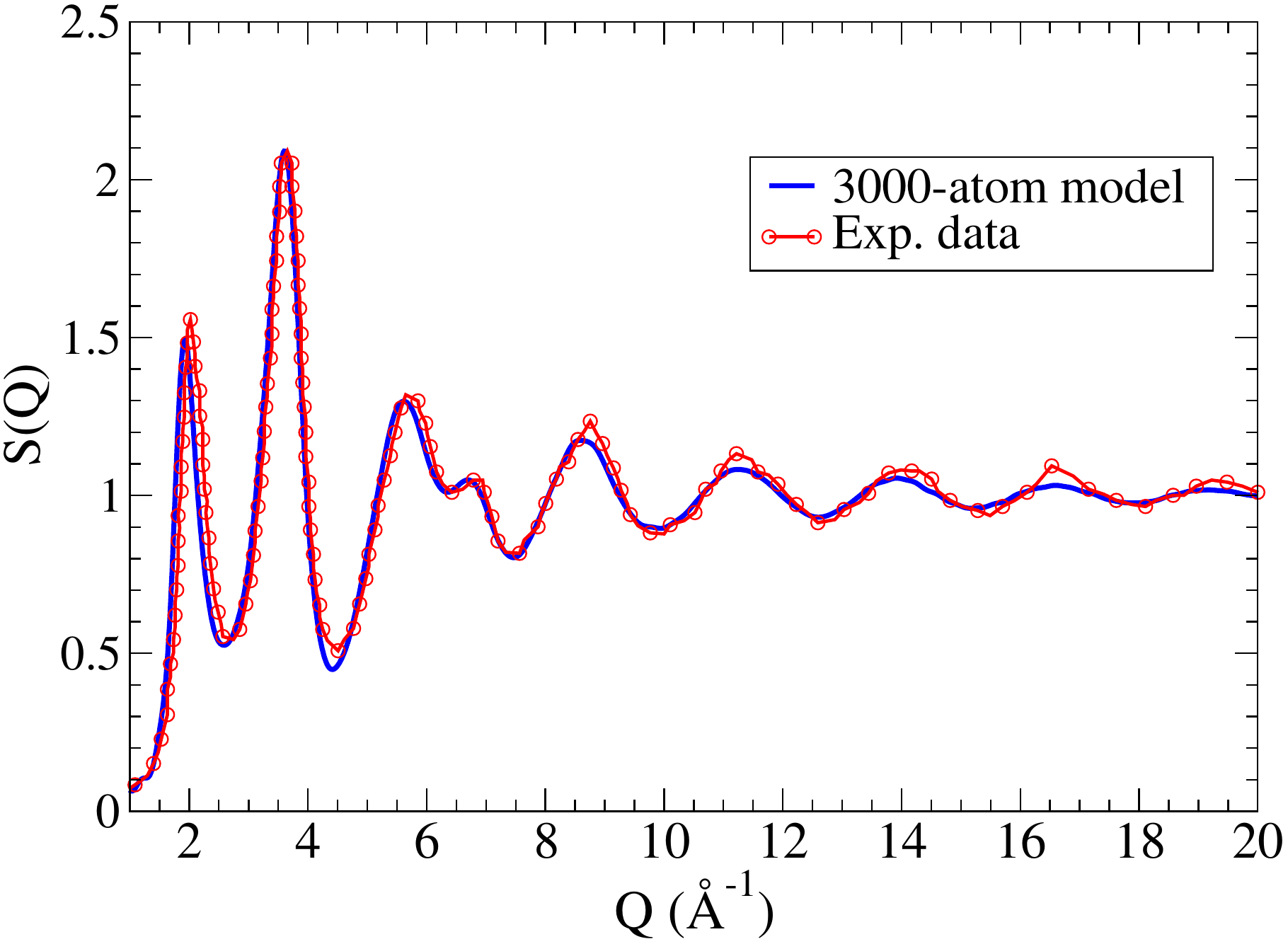}
\caption{ 
The structure factor of {\asi} obtained from simulations 
and experiments. The simulated data are from 3000-atom 
WWW models of density 2.28 g.cm$^{-3}$, whereas the 
experimental data correspond to those from Ref.\,\cite{Laaziri:1999}. 
The simulated data are obtained by averaging over 10 
models from independent runs. 
} 
\label{fig1}
\end{figure}

\begin{figure}[t!]
\centering
\includegraphics[width=.4\textwidth]{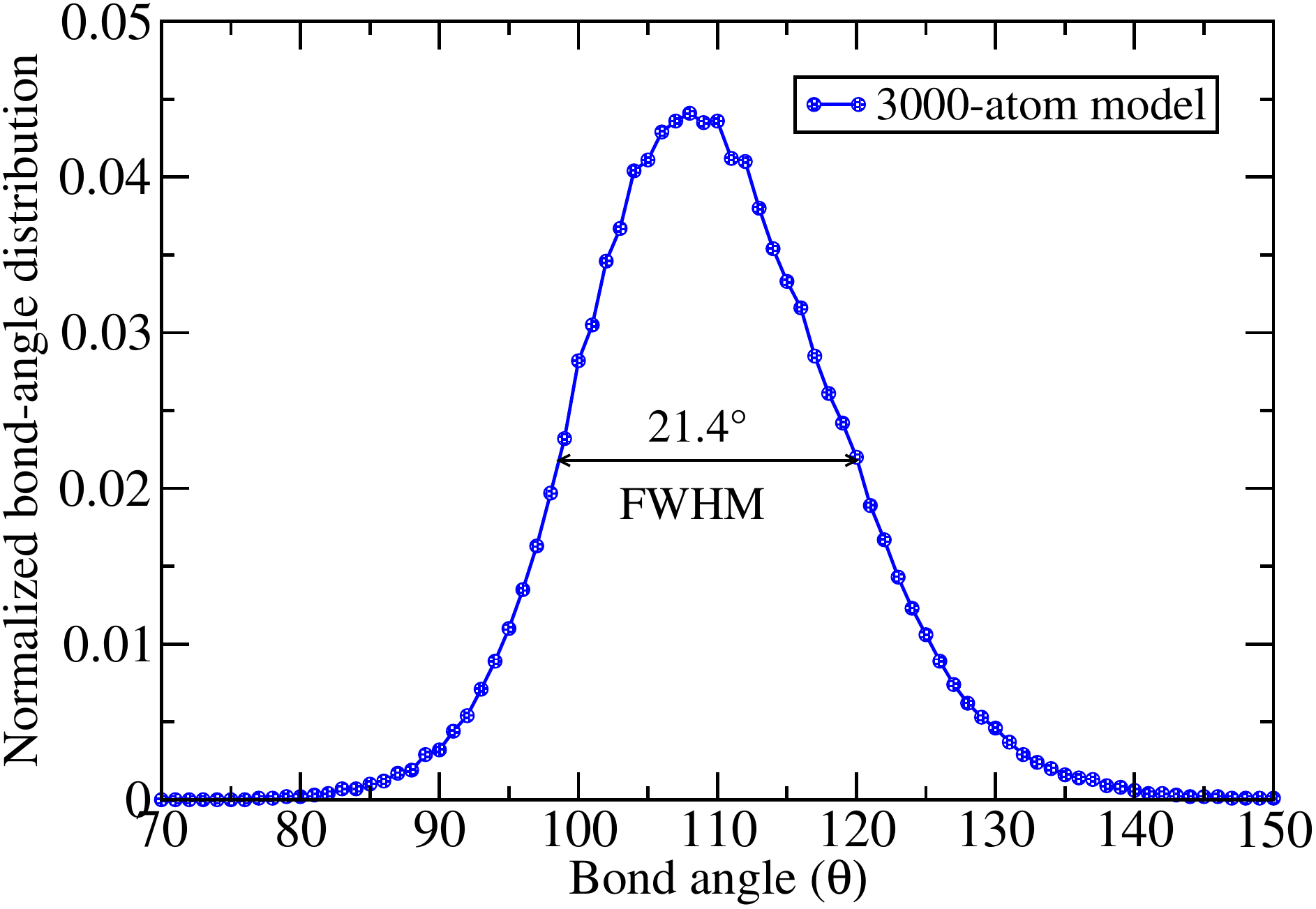}
\caption{ 
The normalized bond-angle distribution for {\asi}, 
obtained for 3000-atom WWW models. The full width 
at half maximum (FWHM) corresponds to a value 
of 21.4\degree. The distribution is obtained by 
averaging over 10 independent models. 
} 
\label{fig2}
\end{figure}

Equation \eqref{E1} suggests that the shape 
of the FSDP can be fully determined via the 
Fourier (sine) transformation of the reduced 
pair-correlation function $G(r)$, provided 
that $G(r) \to 0$ as $r \to R_c$.  Since the 
shape of the FSDP is primarily determined by 
the structure factor in the vicinity of 
$Q_0 \approx$ 2 {\AA}$^{-1}$, it is apparent that 
one requires sufficiently large models of {\asi}, 
in order to satisfy the condition above, for an 
accurate determination of the FSDP.  
To this end, we first validate the structural models of 
{\asi}, obtained from the modified WWW method. 
Since the latter is a well-established method, 
we restrict ourselves to the 
pair-correlation function (PCF), the bond-angle 
distribution (BD), and the coordination number 
(CN) of Si atoms in the network. It has 
been shown elsewhere~\cite{Limbu:2020} that the 
knowledge of the PCF, the BD, and the CN of the atoms are 
sufficient to establish whether a structural 
model can produce the correct electronic 
and vibrational properties of {\asi} or not. 
The full structure factor and the normalized 
bond-angle distribution, obtained from a set 
of 3000-atom models of {\asi}, are plotted in 
Figs.\,\ref{fig1} and \ref{fig2}, respectively. 
For the purpose of configurational averaging, 
the results were averaged over 10 independent 
models of an identical size. The simulated 
values of $S(Q)$ in Fig.\,\ref{fig1} can be 
seen to agree well with the corresponding 
experimental data reported in Ref.\,\cite{Laaziri:1999}. 
Likewise, the full width at half maximum (FWHM) 
of the bond-angle distribution in Fig.~\ref{fig2}, 
about 21.4{\degree}, matches with the observed 
value of 18{\degree}--24{\degree} obtained from 
the Raman ``optic peak" measurements~\cite{Beeman:1985}. 
The FWHM of the bond-angle distribution for the 
WWW models is also found to be consistent with 
those obtained from high-quality molecular-dynamics 
simulations~\cite{JCP2018,Deringer:2018}, 
and data-driven information-based 
approaches~\cite{Limbu:2020,Limbu:2018}, 
developed in recent years. A further characterization 
of the models is possible by examining the statistics 
of the CN of Si atoms, the dihedral-angle distribution, 
and the presence of various irreducible rings in 
the amorphous structures. 
However, since the WWW models have been extensively 
studied and validated in the literature, we will 
not linger over the validation issue and get 
back to the central topic of this paper by listing 
the coordination-number statistics of the 
atoms and some key structural properties of the 
WWW models in Table \ref{tab1}. The corresponding 
results for the DFT-relaxed models are provided 
in Table \ref{tab2}.

\begin{table}[ht!]
\caption{
Structural properties of {\asi} models before DFT relaxation. 
The number of $i$-fold-coordinated atoms (in percent) in the 
network is indicated as $n_i$.  Bond lengths and bond 
angles/widths are expressed in {\AA} and degree, respectively. 
The results are obtained by averaging over 10 configurations 
using a nearest-neighbor cutoff value of 2.8 {\AA}. 
}
\vskip0.2cm
\centering
\begin{ruledtabular}
\begin{tabular}{c|cc|ccc|c}
 \multicolumn{1}{c|} {Size}& \multicolumn{2}{c|} {Bond angle \& width} &  \multicolumn{3}{c|} {Coordination number}& \multicolumn{1}{c}{Bond length} \\
 \cline{1-7}
 $N$ &  $\langle \theta \rangle$ & $\Delta \theta$ & $n_{4}$ &$n_{3} $ & $n_{5} $&$\langle r \rangle$   \\
 \hline
 216   &109.25  & 9.11 & 100   &   0 &  0 & 2.35  \\
 300   &109.25  & 9.32 & 100   &   0 &  0 & 2.35  \\
 512   &109.26  & 9.41 & 100   &   0 &  0 & 2.35 \\
 1000  &109.27  & 9.16 & 100   &   0 &  0 & 2.35  \\
 2000  &109.27  & 9.31 & 99.95 &   0 &  0.05 & 2.35 \\
 3000  &109.26  & 9.39 & 99.94 &   0 &  0.06 & 2.35  \\
 4096  &109.26  & 9.26 & 99.95 &   0 &  0.05 & 2.35  \\
 5000  &109.27  & 9.31 & 99.97 &   0 &  0.03 & 2.35 \\
 6000  &109.26  & 9.39 & 99.96 &   0 &  0.04 & 2.35 
\end{tabular}
\end{ruledtabular}
\label{tab1}
\end{table}

\begin{table}[ht]
\caption{
Structural properties of DFT-relaxed models of $\it{a}$-Si.  The 
total number of $i$-fold-coordinated atoms (in percent) present 
in the relaxed networks is indicated as $n_i$. Average bong 
lengths and bond angles/widths are expressed in {\AA} and degree, 
respectively. Asterisks indicate the use of single-zeta basis 
functions and the non-self-consistent Harris-functional 
approximation for relaxation of large models. 
}
\vskip0.5cm
\centering
\begin{ruledtabular}
\begin{tabular}{ c| cc | c c c  |c  }
 \multicolumn{1}{c|} {Model size}& \multicolumn{2}{c|} {Bond angle \& width} &  \multicolumn{3}{c|} {Coordination number}& \multicolumn{1}{c}{Bond length} \\
 \cline{1-7}
 $N$ &  $\langle \theta \rangle$ & $\Delta \theta$ & $n_{4}$ &$n_{3} $ & $n_{5} $&$\langle r \rangle$   \\
 \hline
 216 &\: 109.11 \:   & 10.14 &100 &   0 &  0 & 2.36  \\
 300 &\: 109.15 \:   & 10.22 & 100 &   0 &  0 & 2.36  \\
 512 & \:  109.13\:  & 10.45 & 100 &   0 &  0 & 2.36 \\
 1000 &\: 109.15\:   & 10.14 & 100 &   0 &  0 & 2.36  \\
 2000 &\: 109.14\:   & 10.3 &\: 99.95 &   0 &  0.05 & 2.36 \\
 3000 &\: 109.13\:   & 10.4 &\: 99.96 &\:  0.01 &  0.03 & 2.36  \\
 4096$^*$ &\: 109.02\:  & 10.82 &\: 99.95 & \:  0.01 &  0.04 & 2.37 \\
 5000$^*$ &\: 109.01\:  & 10.9 &\: 99.94 &\:  0.01 &  0.05 & 2.37 \\
 6000$^*$ & \: 109.51\:  & 10.92 &\: 99.95 &\:   0.01 &  0.04 & 2.37  \\
\end{tabular}
\end{ruledtabular}
\label{tab2}
\end{table}

\subsection{Characterization, origin, and the structure of the FSDP in 
amorphous silicon
}
\begin{figure}[t!]
\centering
\includegraphics[width=.4\textwidth]{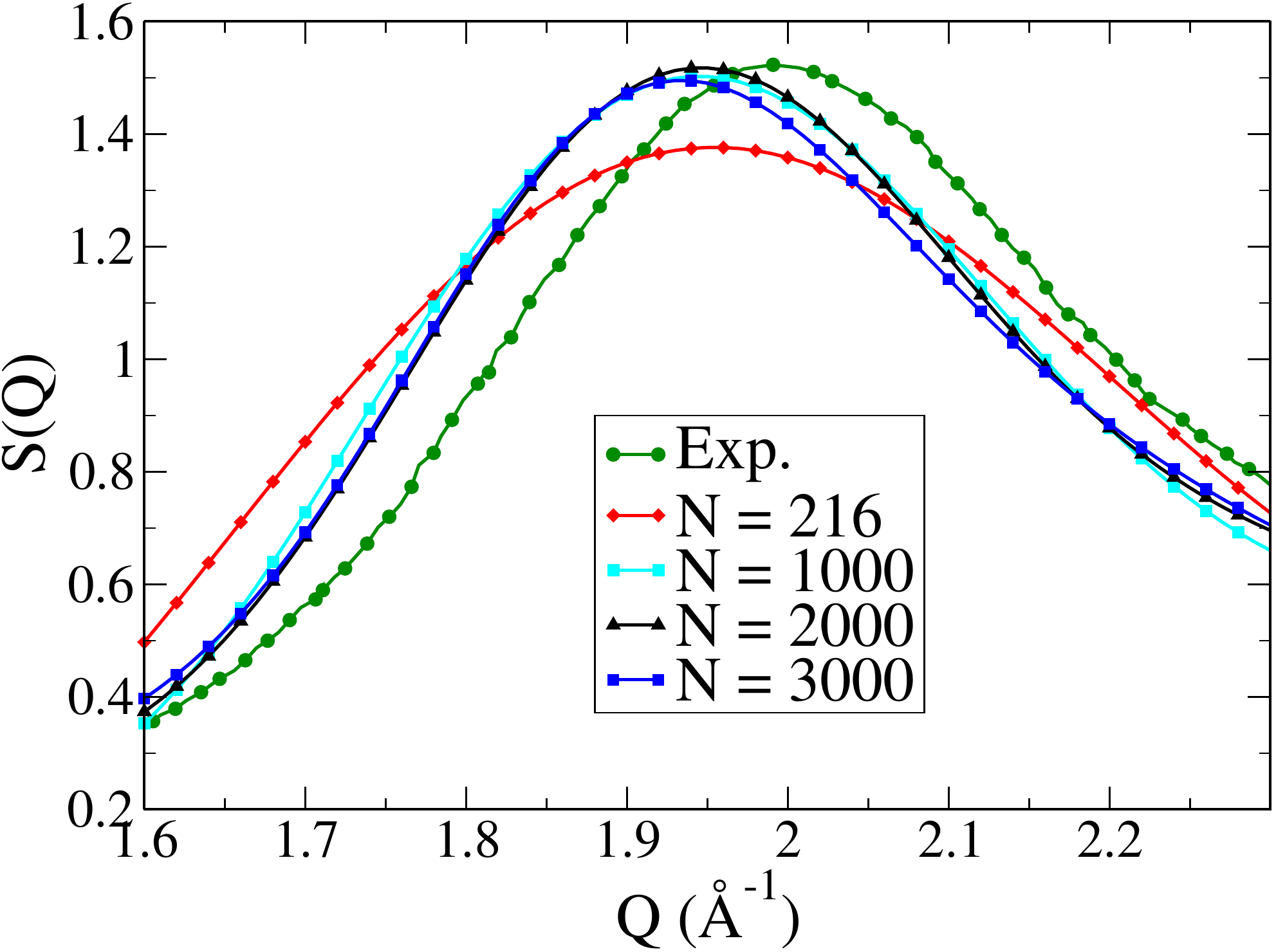}
\caption{ 
The structure factor of {\asi} in the vicinity of 
the FSDP from simulations and experiments. 
Experimental data (\textcolor{ForestGreen}{\textbullet}) 
correspond to as-implanted samples from Ref.\,\cite{Xie:2013}, 
whereas simulated data refer to 216-atom 
(\textcolor{red}{{\tiny $\blacklozenge$}}), 
1000-atom (\textcolor{cyan}{{\tiny $\blacksquare$}}), 
2000-atom ($\blacktriangle$), and 3000-atom 
(\textcolor{blue}{{\tiny $\blacksquare$}}) unrelaxed 
WWW models.
} 
\label{fig3}
\end{figure}

\begin{figure}[t!]
\centering
\includegraphics[width=.4\textwidth]{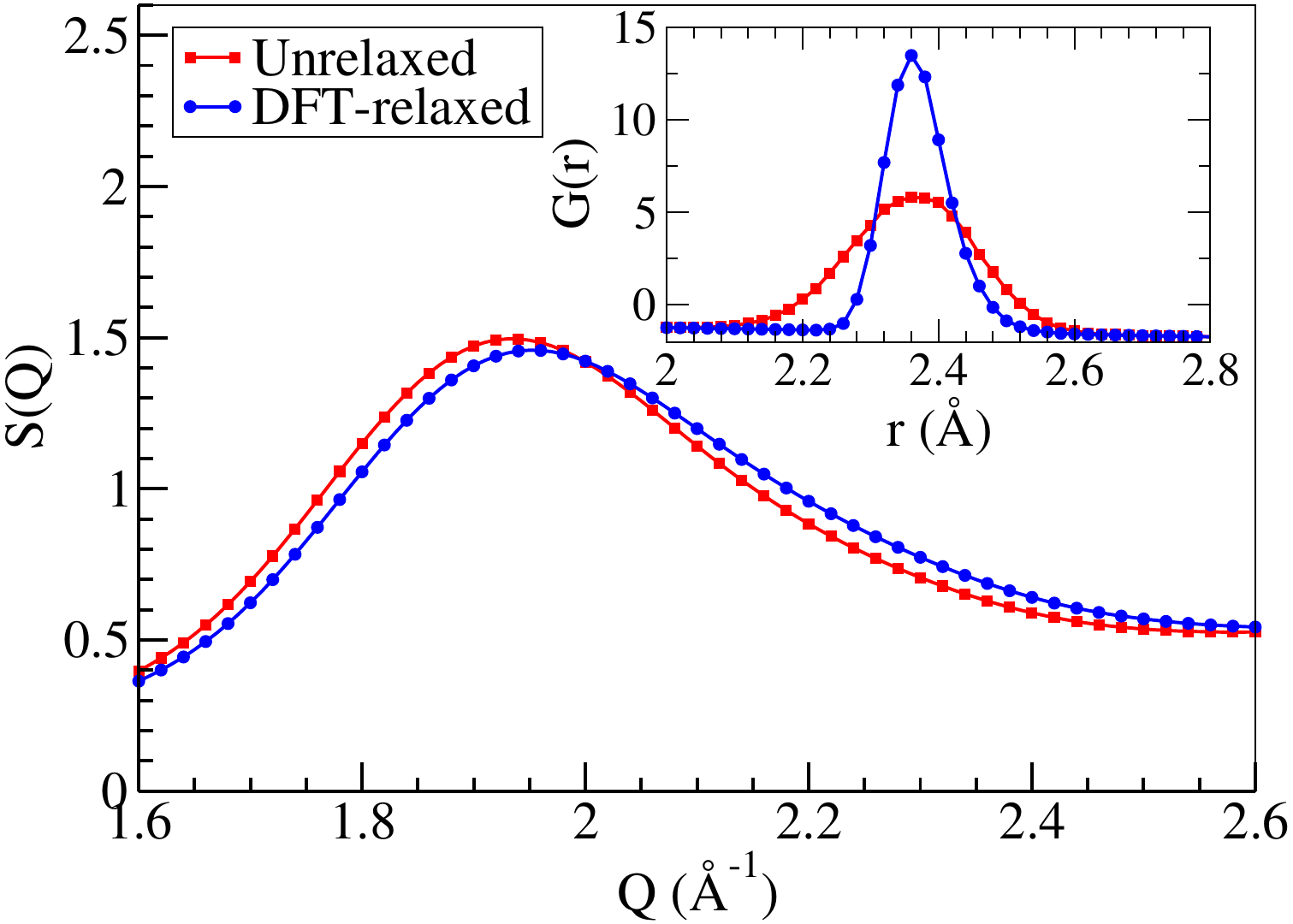}
\caption{ 
Effects of approximate first-principles relaxations 
on the position and the intensity of the FSDP of {\asi} 
for a 3000-atom model before (\textcolor{red}{\tiny{$\blacksquare$}}) 
and after (\textcolor{blue}{\large$\bullet$}) relaxation. A small shift of 
the diffraction peak toward higher values of $Q$ is 
accompanied by a slight reduction of the peak intensity 
in the relaxed model. The corresponding reduced pair-correlation 
functions near the first peak are shown in the inset. 
} 
\label{fig4}
\end{figure}

Figure \ref{fig3} shows the structure factor of {\asi} 
obtained from four different models, of size from 216 
atoms to 3000 atoms, and experiments~\cite{Xie:2013}. 
As before, the simulation data are obtained by averaging 
over 10 independent models for each size, whereas 
the experimental data refer to as-implanted 
samples of {\asi} in Ref.\,\cite{Xie:2013}.  
An examination of Fig.\,\ref{fig3} leads to the 
following observations. Firstly, it is 
apparent that the 216-atom model shows a marked deviation 
from the experimental data near the FSDP, indicating 
noticeable finite-size effects originated from small 
models of linear size of about 16 {\AA}. By 
contrast, the larger models, consisting of 1000 to 3000 
atoms, have produced the peak intensity more accurately. 
Secondly, all the models consistently underestimate the 
position of the experimental FSDP~\cite{Xie:2013} at 
$Q_0$=1.99 {\AA}$^{-1}$, by an amount of 0.045 {\AA}$^{-1}$. 
One can surmise a number of possible reasons 
for this discrepancy. These include the 
inadequacy of the classical potentials, the uncertainty 
of the actual density of the {\asi} sample(s) 
used in experiments, and a possible 
sample-to-sample dependence of the experimental 
results. The last point can be appreciated by noting 
that the experimental value of $Q_0$,  
for as-implanted samples of {\asi}, reported in 
Refs.~\cite{Fortner:1989}, \cite{Xie:2013}, and 
\cite{Laaziri:1999} differ from each other by about 
0.07 {\AA}$^{-1}$ (see Fig.~\ref{fig9}). Finally, 
a first-principles total-energy relaxation of the 
models, using the density-functional code {\sc 
Siesta}~\cite{Soler:2002}, somewhat remedies this 
issue at the expense of the reduction of the peak intensity. 
This is illustrated in Fig.\,\ref{fig4}, 
where we have plotted both the reduced pair-correlation 
function (see inset) and the configurational-average 
structure factor from ten 3000-atom models before and 
after total-energy relaxation. The increase of the 
peak height of $G(r)$ upon relaxation 
is not surprising in view of the fact that first-principles 
relaxations minimized the total energy of the system 
by reducing the bond-length disorder at the expense of 
a minor increase of the bond-angle disorder. The latter is 
reflected in the root-mean-square (RMS) width, 
$\Delta\theta$, of the bond-angle distribution before and 
after relaxation in Tables \ref{tab1} and \ref{tab2}, 
respectively. By contrast, the shape of the structure 
factor near the FSDP remains more or less the same 
after relaxation, except for a small shift of the 
FSDP toward the higher values of $Q$. 

\begin{figure}[tb!]
\centering
\includegraphics[width=0.42\textwidth]{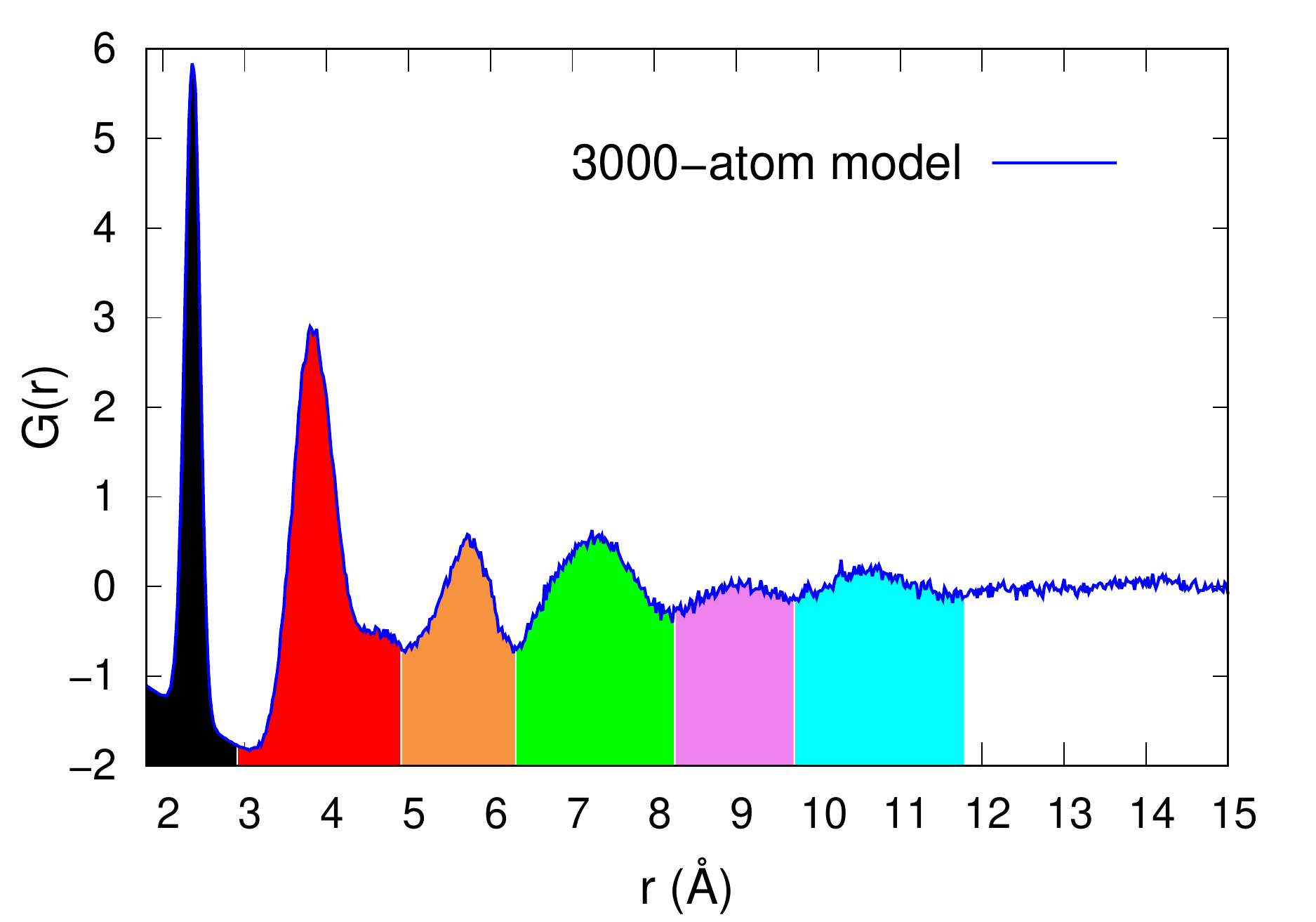}
\caption{
The reduced pair-correlation function, $G(r)$, of {\asi} 
obtained from configurational averaging of ten 3000-atom 
WWW models.  The presence of the first six 
radial shells, which extend up to a distance of $\approx$ 12 {\AA}, 
is highlighted in different colors.  
}
\label{fig5}
\end{figure}

Having addressed the overall shape of the structure 
factor and the FSDP for a number of models of varying 
sizes, we now examine the origin 
of the FSDP in terms of the real-space structure of {\asi} 
networks. While it is well-understood that the FSDP 
in {\asi} arises from the medium-range order in the 
network, which entails a length scale of a few 
to several angstroms, a quantitative characterization 
of the contribution from different radial shells 
is still missing in the literature. 
We address this aspect of the problem by examining 
the role of radial atomic correlations in forming 
the FSDP, via the Fourier transform of the reduced PCF, 
and provide a quantitative measure of the contributions 
that originate from the increasingly distant radial 
shells in the amorphous environment of silicon. 
This can be achieved by writing,
\[ 
S(Q) = 1 + F(Q) = 1 + \sum_{i=1}^n F_i(Q; R'_i, R'_{i+1}),
\] 
where, 
\be 
F_i(Q; R'_i, R'_{i+1}) = \dfrac{1}{Q}\int_{R'_i}^{R'_{i+1}} G(r)\, \sin(Qr)\, dr. 
\label{E3}
\ee
\noindent
In Eq.\,(\ref{E3}), $F_i(Q; R'_i, R'_{i+1})$ is the contribution 
to $F(Q)$ from the reduced PCF, $G(r)$, at distances between 
$R'_i$ and $R'_{i+1}$.  The contribution from a given radial 
shell can be obtained by a suitable choice of $R'_i$ and 
$R'_{i+1}$, where $R'_{i+1} > R'_i$, and an appropriate set 
\{$R'_1, \ldots, R'_n$\} covers the entire radial (integration) 
range to obtain the full $F(Q)$.  For example, a choice of 
$R'_1$ = 0 {\AA} and $R'_2$ = 2.8 {\AA} yields $F_1(Q; R'_1, R'_2)$, 
and $R'_2$ = 2.8 {\AA} and $R'_3$ = 4.9 {\AA} provides $F_2(Q; R'_2, R'_3)$. 
The origin of the FSDP and the principal peak can be studied 
by computing various $F_i(Q)$ in the vicinity of 2 {\AA}$^{-1}$ 
and 3.6 {\AA}$^{-1}$, respectively.  The appropriate values of 
$R'_i$ for different radial shells can be obtained by inspecting 
the reduced PCF of {\asi}. This is illustrated in Fig.\,\ref{fig5}, 
by plotting the configurational-average $G(r)$ obtained from a 
set of ten 3000-atom models. We should emphasize that the radial 
shells correspond to the radial regions between two neighboring 
minima in the reduced PCF. Except for the first radial shell, the 
radial regions, defined by a pair of consecutive minima in $G(r)$, 
are not necessarily identical to the corresponding atomic-coordination 
shells due to the overlap of the atomic distribution from 
different coordination shells.

\begin{figure}[t!]
\centering
\includegraphics[width=.4\textwidth]{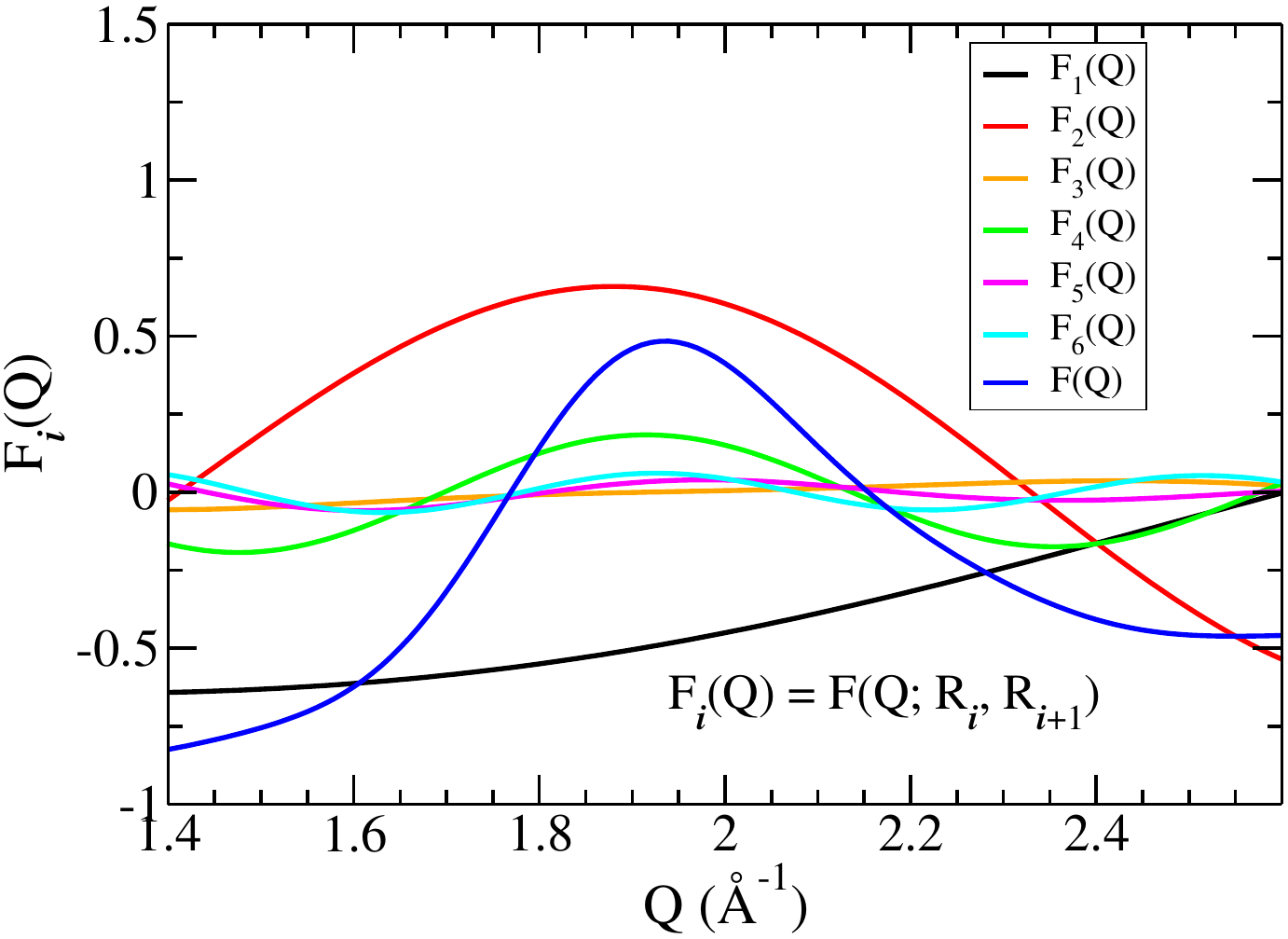}
\caption{
The contribution to the FSDP, $F_i(Q)$, near $Q_0$, 
originating from the first six radial shells and 
the total $F(Q)$ (blue).  The results correspond 
to 3000-atom WWW models, and are averaged over 
ten configurations. The color of the plots 
corresponds to the color of the radial shells 
in Fig.~\ref{fig5}. 
} 
\label{fig6}
\end{figure}

\begin{figure}[t!]
\centering
\includegraphics[width=0.42\textwidth]{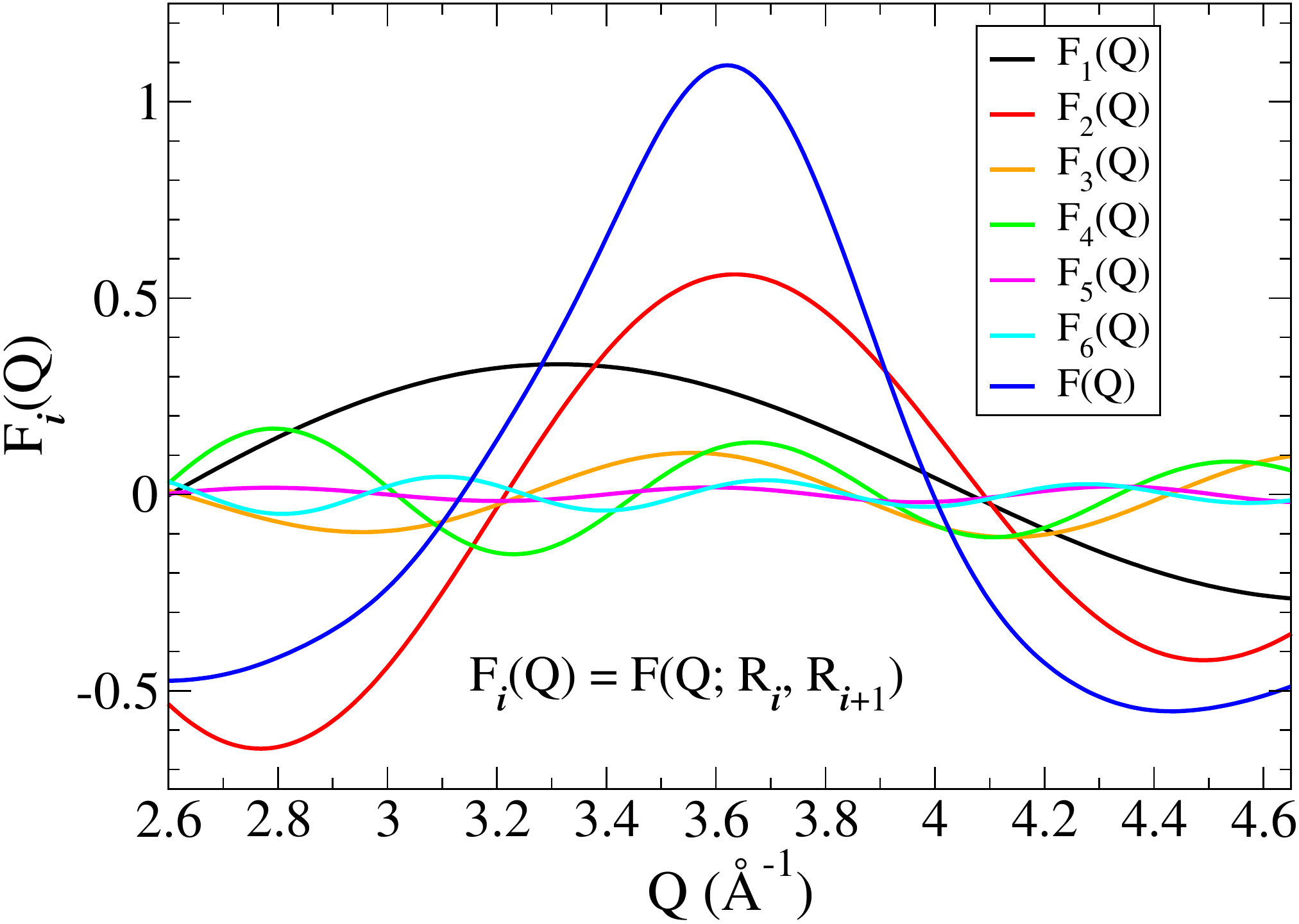}
\caption{
The contribution, $F_i(Q)$, to the principal peak 
at $Q_1$ = 3.6 {\AA}$^{-1}$ from the first six 
radial shells of the reduced 
pair-correlation function. The total $F(Q)$ 
is shown in blue color for comparison.  The 
results are obtained via configurational 
averaging of data from ten 3000-atom models.  
}
\label{fig7}
\end{figure}

Figure \ref{fig6} shows the contribution to $F(Q)$ in the 
vicinity of 2 {\AA}$^{-1}$ from the first six radial 
shells. The plots for different radial shells are indicated 
by the corresponding shell color as depicted in Fig.~\ref{fig5}. 
It is evident that the chief contribution to the FSDP 
comes from $F_2(Q)$, which is followed by $F_4(Q)$ and 
$F_6(Q)$ in the descending order of magnitude. $F_2(Q)$ 
and $F_4(Q)$ play a crucial role in determining both 
the intensity and the position of the FSDP, while $F_3(Q)$ 
and $F_5(Q)$ contribute very little to none. By contrast, 
$F_1(Q)$ monotonically changes in the vicinity of the FSDP 
and thus contributes to the intensity (and the shape) of 
the FSDP near $Q_0$ to some degree but does not play 
any noticeable role in determining the position of $Q_0$.  
It is therefore 
apparent that the position of the FSDP in {\asi} is 
primarily determined by the information from the second 
radial shell, followed by the fourth and sixth radial 
shells, whereas the rest of the distant radial shells 
provide small perturbative corrections. The enumeration of the radial 
shell-by-shell contribution to $F(Q)$ is a significant 
result to our knowledge, which cannot be quantified from 
a phenomenological understanding of the Fourier transform 
of $G(r)$ in Eq.\,\eqref{E1}. A similar analysis reveals 
that the contribution to the principal peak at 3.6 {\AA}$^{-1}$ 
mostly arises from $F_2(Q)$, $F_1(Q)$, $F_4(Q)$, and 
$F_3(Q)$, in the decreasing order of magnitude.  Once 
again, $F_1(Q)$ is found to provide a positive but 
monotonically decreasing contribution with increasing $Q$ 
in the vicinity of the principal peak. Thus, the peak at 
3.6 {\AAI} is principally contributed by the first four 
radial shells in the reduced PCF.  This observation 
amply justifies the use of the term `principal peak' 
to describe the peak at 3.6 {\AAI} in the structure 
factor of {\asi}.  Figure \ref{fig7} shows the results 
for the principal peak using the same color code as in 
Fig.~\ref{fig5}. 

\subsection{Relation between peaks in {\it S(Q)} and {\it G(r)}}
The results presented in the preceding section on the 
basis of the partitioning of $F(Q)$ reveal that the 
information from the second and fourth radial shells 
largely determine the structure, i.e., the 
position, intensity, and width, of the FSDP in {\asi}. 
We now provide a physical interpretation of 
the numerical results and demonstrate that 
the emergence of the first two peaks in $S(Q)$, near 
2 {\AA}$^{-1}$ and 3.6\,{\AA}$^{-1}$, respectively, 
can be deduced simply from the knowledge of the reduced PCF 
and the behavior of the integral, involving the 
sinc(x) (i.e., $\sin x/x$) function, which defines 
the structure factor.  Noting that the structure 
factor can be written as, 
\bea
S(Q) &=& 1 + F(Q) = 1 + \frac{1}{Q}\int_0^{R_c} G(r) \sin(Q\,r)\, dr \notag \\
     &=& 1 + \int_0^{R_c} r\,G(r) \left[\dfrac{\sin(Q\,r)}{Qr}\right] \, dr,
\eea
it is elementary that the peaks in $F(Q)$ (and hence $S(Q)$) 
should appear approximately for those values of $Qr$ 
for which both $\sin(Qr)/Qr$ and $rG(r)$ are maximum. Here, 
the $r$ values in $Qr$ are given by the maxima of $rG(r)$.  
Since the maxima of $\sin(Qr)/Qr$ and $\sin(Qr)$ are very 
close to each other~\cite{sinc} for $Qr > 0$, and the 
maxima of $G(r)$ and $rG(r)$ practically coincide, one 
may use the maxima of $\sin(Qr)$ and $G(r)$ in determining the 
approximate location of the first two peaks in $S(Q)$. 
This implies $Qr$ must satisfy, $\sin(Qr)$ = 1, or 
$Qr = (4m+1)\pi/2$, where $m$=0, 1, 2, \ldots etc.  
Since the first two maxima of $G(r)$ are given by $r_1 \approx$ 2.35 
{\AA} and $r_2 \approx$ 3.8 {\AA}, respectively, and $m$ = 0 does not admit 
a physical solution~\cite{sinc}, the first major contribution to 
the $F(Q)$ comes from the second radial shell for $r_2$ = 3.8 {\AA} 
and $m = 1$.  This gives, $Q_0 = 5\pi/(2\times r_2)$ 
= 2.07 {\AA}$^{-1}$.  Likewise, the next contribution 
comes from, for $m = 2$, the fourth radial shell 
with a peak at $r_4 \approx$ 7.24 {\AA} in $G(r)$. This yields, 
$Q_0 = 9\pi/(2 \times r_4)$ = 1.95 {\AA}$^{-1}$. 
A similar analysis shows that the principal peak ($Q_1$) 
gets its share from the first radial shell, for $m=1$, 
at $Q_1 = 5\pi/(2\times r_1)$ = 3.34 {\AA}$^{-1}$, 
which is followed by the second radial shell, for 
$m=2$, at $Q_1 = 9\pi/(2\times r_2)$ = 3.72 {\AA}$^{-1}$, 
the fourth radial shell, for $m=4$, at $Q_1 
= 17\pi/(2\times r_4)$ = 3.69 {\AA}$^{-1}$, and 
the third radial shell, for $m=3$ and $r_3= 5.72$, 
at $Q_1 = 13\pi/(2\times r_3)$ = 3.57 {\AA}$^{-1}$. 
The exact position of a peak in $S(Q)$ is 
determined by the sum of the contribution 
from the relevant radial shells, which 
introduce a minor deviation from the individual estimate 
above due to the approximate nature of our calculations. 
Table \ref{tab3} presents a summary of the results 
obtained from the reasoning above. The estimated 
position of the peaks in $F_i(Q)$, for $i$=1 to 6, 
is listed in the Table. The first column, 
shown in light gray shading, corresponds to the maxima 
($r_i$) of $G(r)$ up to a 
radial distance of 11 {\AA}, whereas the second row, 
indicated by dark gray cells, lists the values 
of $Q\,r = (4m+1)\pi/2$ for $m$=1 to 6. The remaining six rows, 
between columns 1 and 8, indicate the peak positions 
in $F_i(Q)$ that are obtained by dividing the $Qr$ values by 
the corresponding $r_i$ value from the first column. 
The estimated positions of the FSDP and the principal 
peak for a number of combination of $(r_i, m)$ 
are shown in Table \ref{tab3} by green and red 
colors, respectively. 

Conversely, assuming that the FSDP is located at 
$Q_0 \approx$ 2 {\AA}$^{-1}$, one arrives at the 
conclusion, by dint of our logic, that the 
contribution to the FSDP should come from 
$r = (4m+1)\pi/(2 \times Q_0)= \pi/4, 5\pi/4,
9\pi/4$, and $13\pi/4$, etc., for $m$ = 
0 to 3. The first value of $r$, for $m$ = 0, 
does not provide a physical solution but the 
remaining values at 3.93 {\AA}, 7.07 {\AA}, 
and 10.21 {\AA} approximately correspond to the 
second peak, the fourth peak, and the sixth 
peak of $G(r)$ (cf.\, Fig.\,\ref{fig5}).  A 
similar analysis can be done for the principal peak.  
The argument presented here suffices to explain 
why the information from the distant radial shells, 
for $r \ge 15$ {\AA}, cannot contribute 
significantly in the formation of the FSDP. At large 
radial distances, when the reduced PCF rapidly vanishes 
and the concomitant numerical noises in $G(r)$ 
become increasingly stronger, $\sin(Qr)/Qr$ cannot 
find, or sample, suitable values of $r$ with a large 
$G(r)$, for $Q$ values near the FSDP, to satisfy 
the condition above. This leads to small $F_i(Q)$ 
for the distant radial shells. We have verified 
that the analysis presented here is consistent 
with the results from numerical calculations of 
$F_i(Q)$.

\begin{table}[t!]
\caption{
Estimated values of the peak positions, $Q$, in $F_i(Q)$, 
obtained from $Q\,r = (4m+1)\pi/2$ (dark gray cells in 
the second row for $m$=1 to 6) and the maxima of $G(r)$ 
(first column) in {\AA}. The positions of the FSDP and 
the principal peak (PP) in $F_i$s are indicated by green 
and red colors, respectively. The radial shells that 
contribute to the FSDP and the PP can be directly read 
off the first column.
}
\vskip0.5cm
\centering
\begin{tabular}{a|c|c|c|c|c|c|c}
\hline
Maxima     & {1}  & {2}    & {3}    & {4}   & {5}    & {6}  & $\leftarrow$ $m$ \\
\cline{2-8} 
\rowcolor{dgray}
\cellcolor{lightgray} of $G(r)$ 
&7.854 & 14.137 & 20.42 & 26.704 & 32.987 & 39.27 & $\leftarrow$ $Qr$\\
\hline
\cline{2-8}
2.35 & \cellcolor{salmon}3.34 & \cellcolor{white} 6.02 &  \cellcolor{white}8.69 &  11.36 & 14.04 & 16.71 & Peaks in $F_{1}$ \\
\hline
3.8 & \cellcolor{green1} 2.07 & \cellcolor{salmon}3.72 & \cellcolor{white} 5.37 & 7.03 &  \cellcolor{white}8.68 & 10.33 & Peaks in $F_{2}$  \\
\hline
5.72 & 1.37 & 2.47 & \cellcolor{salmon}3.57 & 4.67 &\cellcolor{white} 5.77 & 6.87 & Peaks in $F_3$  \\
\hline
7.24 & 1.08 & \cellcolor{green1}1.95 & 2.82 &  \cellcolor{salmon}3.69 & 4.56 & \cellcolor{white} 5.42 & Peaks in $F_4$  \\
\hline
9.16 & 0.86 & 1.54 & 2.23 & 2.92 &  3.6 & 4.29 & Peaks in $F_5$ \\
\hline
10.74& 0.73 & 1.32 & \cellcolor{green1}1.9 & 2.49 & 3.07 &  3.66 & Peaks in $F_6$\\
\hline
\end{tabular}
\label{tab3}
\end{table}
\begin{figure}[h!]
\centering
\includegraphics[width=.42\textwidth]{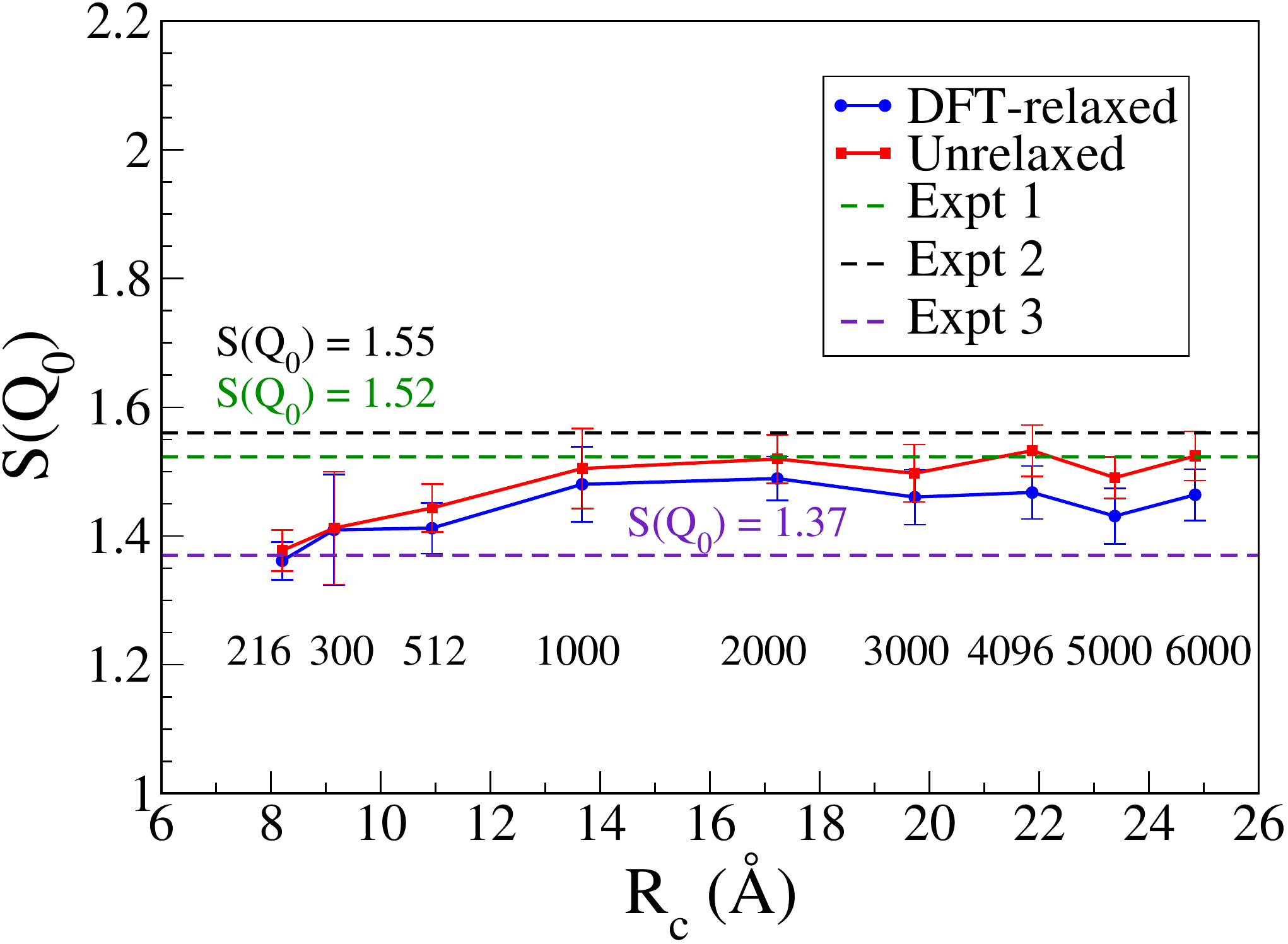}
\caption{
The dependence of the intensity of the FSDP, $S(Q_0)$, with the
radial cutoff distance, $R_{c}$, for a number of models of
different sizes, as indicated in the plot. The experimental 
values of $S(Q_0)$ reported in the literature are shown as 
horizontal dashed lines: 1) $S(Q_0)$=1.52 from 
Ref.~\cite{Xie:2013}; 2) $S(Q_0)$=1.37 from Ref.~\cite{Fortner:1989}; 
and 3) $S(Q_0)$=1.55 from Ref.~\cite{Laaziri:1999}. 
}
\label{fig8}
\end{figure}

The results and discussion presented so far indicate that 
the radial information from the reduced PCF of up to a length 
scale of 15 {\AA} plays a significant role in the 
formation of the FSDP. To further establish this point, we 
now conduct a systematic study of the structure of the FSDP 
in terms of the intensity and the width of the peak. 
The variation of the peak intensity with the size of 
the models is studied by plotting the value of 
$S(Q_0)$ against $R_c$ for a number of DFT-relaxed/unrelaxed 
models, consisting of 216 atoms to 6000 atoms. 
Since $R_c$ is given by the half of the linear size 
of the models, Fig.\,\ref{fig8} essentially shows 
the dependence of $S(Q_0)$ on the radial pair 
correlations up to a distance of $R_c$, through 
the Fourier transform of $G(r)$. 
It is clear from the plots (in Fig.~\ref{fig8}) that the intensity 
of the FSDP for both the relaxed and unrelaxed models 
varies considerably until $R_c$ increases to a value 
of the order of 14 {\AA}. This roughly translates into 
a model of size about 1000 atoms. 
For even larger values of $R_c$,  the peak intensity 
is more or less converged to 1.48 for the unrelaxed 
models but considerable deviations exist for the value 
of DFT-relaxed models from the experimental value 
of $S(Q_0)$ of 1.52 in Ref.~\cite{Xie:2013}. 
The deviation of the peak intensity from the experimental 
value for small models of {\asi} can be readily understood. 
Since $G(r)$ carries considerable real-space information up to a 
radial distance of 15 {\AA}, possibly 20 {\AA} for very 
large models, small models with $R_c$ values less than 15 {\AA} cannot 
accurately produce the peak position using Eq.~\eqref{E1}. 
On the other hand, the peak intensity for the DFT-relaxed 
models deviates noticeably (about 0.2--12\%) from their 
unrelaxed counterpart and the experimental value for 
as-implanted samples in Refs.~\cite{Xie:2013},~\cite{Fortner:1989}, 
and~\cite{Laaziri:1999}. This apparent deviation for the 
bigger models is not particularly unusual and it can be 
attributed, at least partly, to: 1) the use of approximate 
total-energy calculations in the relaxation of large models, 
via the non-self-consistent Harris-functional approach 
using minimal single-zeta basis functions; 2) the intrinsic difficulties 
associated with quantum-mechanical relaxations of large 
models; and 3) the sample dependence of experimental results, 
showing a considerable difference in the value of 
$S(Q_0)$ for as-implanted samples in Fig.~\ref{fig8}, 
which is as high as 0.18 from one experiment to another. 
Thus, the results obtained in this study are well within 
the range of the experimental values reported in the 
literature~\cite{Fortner:1989,Xie:2013,Laaziri:1999}. 

\begin{figure}[t!]
\centering
\includegraphics[width=.4\textwidth]{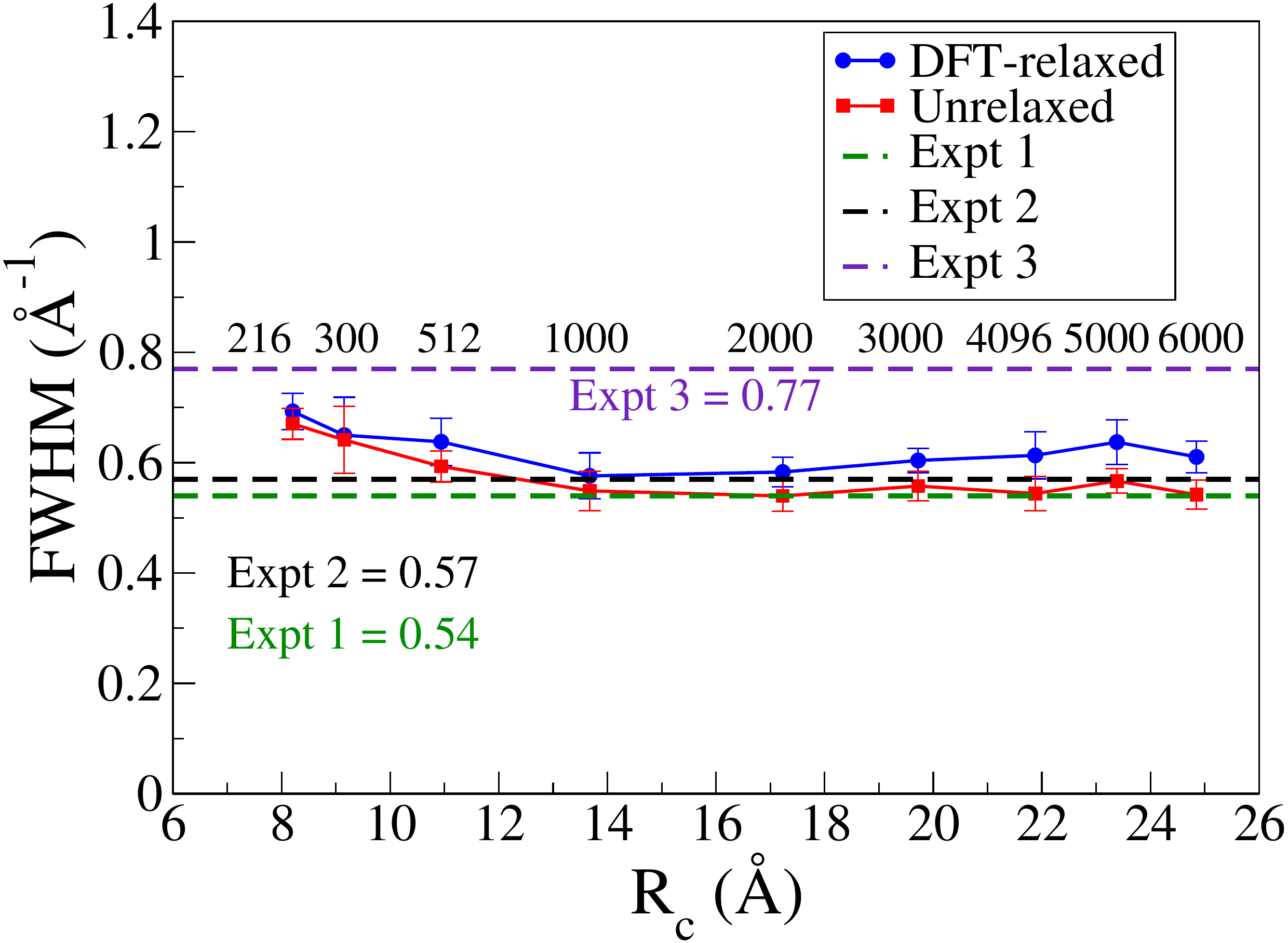}
\caption{
The full width at half maximum (FWHM) of the FSDP at 
$Q_0$ for a number of models before 
(\textcolor{red}{\tiny{$\blacksquare$}}) and after 
(\textcolor{blue}{\large $\bullet$}) DFT relaxations. 
The horizontal dashed lines indicate the experimental values 
of 0.54 {\AA}$^{-1}$ (green), 0.57 {\AA}$^{-1}$ (black), 
and  0.77 {\AA}$^{-1}$ (indigo) for as-implanted 
samples of {\asi} from Refs.\,\cite{Xie:2013}, 
\cite{Laaziri:1999}, and \cite{Fortner:1989}, 
respectively. 
} 
\label{fig9}
\end{figure}

\begin{figure}[ht!]
\centering
\includegraphics[width=.4\textwidth]{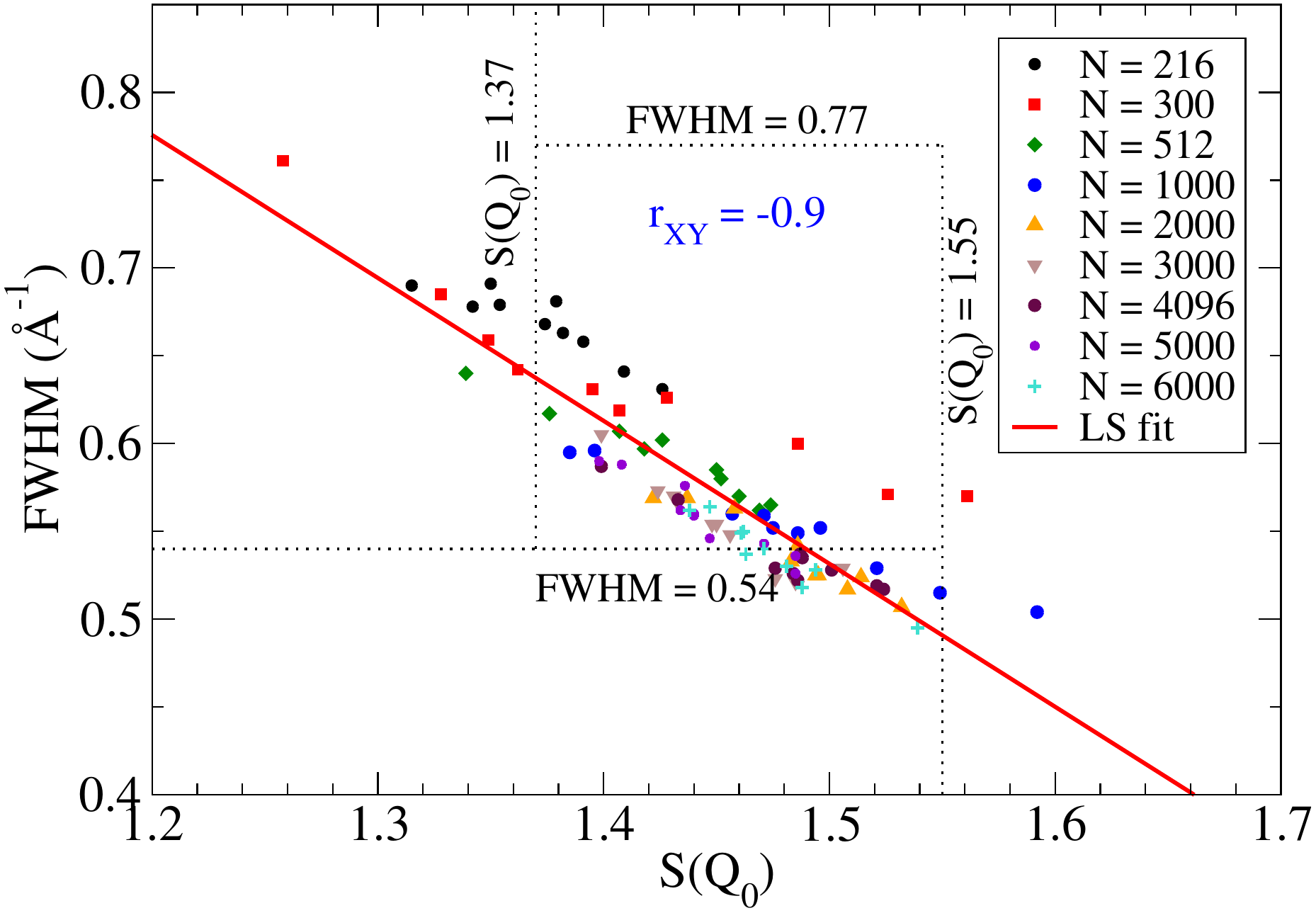}
\caption{
A scattered plot showing the presence a clear 
correlation between the FWHM and $S(Q_0)$ of 
the FSDP for a number of models of varying system 
sizes.  The solid (red) line corresponds to the 
linear least-square (LS) fit of the data, whereas 
$r_{XY} = -0.9$ indicates the Pearson correlation 
coefficient for the data sets. The horizontal 
and vertical dotted lines indicate the experimental 
values of FWHM and $S(Q_0)$, respectively, obtained 
for as-implanted samples of {\asi}.
} 
\label{extra}
\end{figure}
\begin{figure}[t!]
\centering
\includegraphics[width=.4\textwidth]{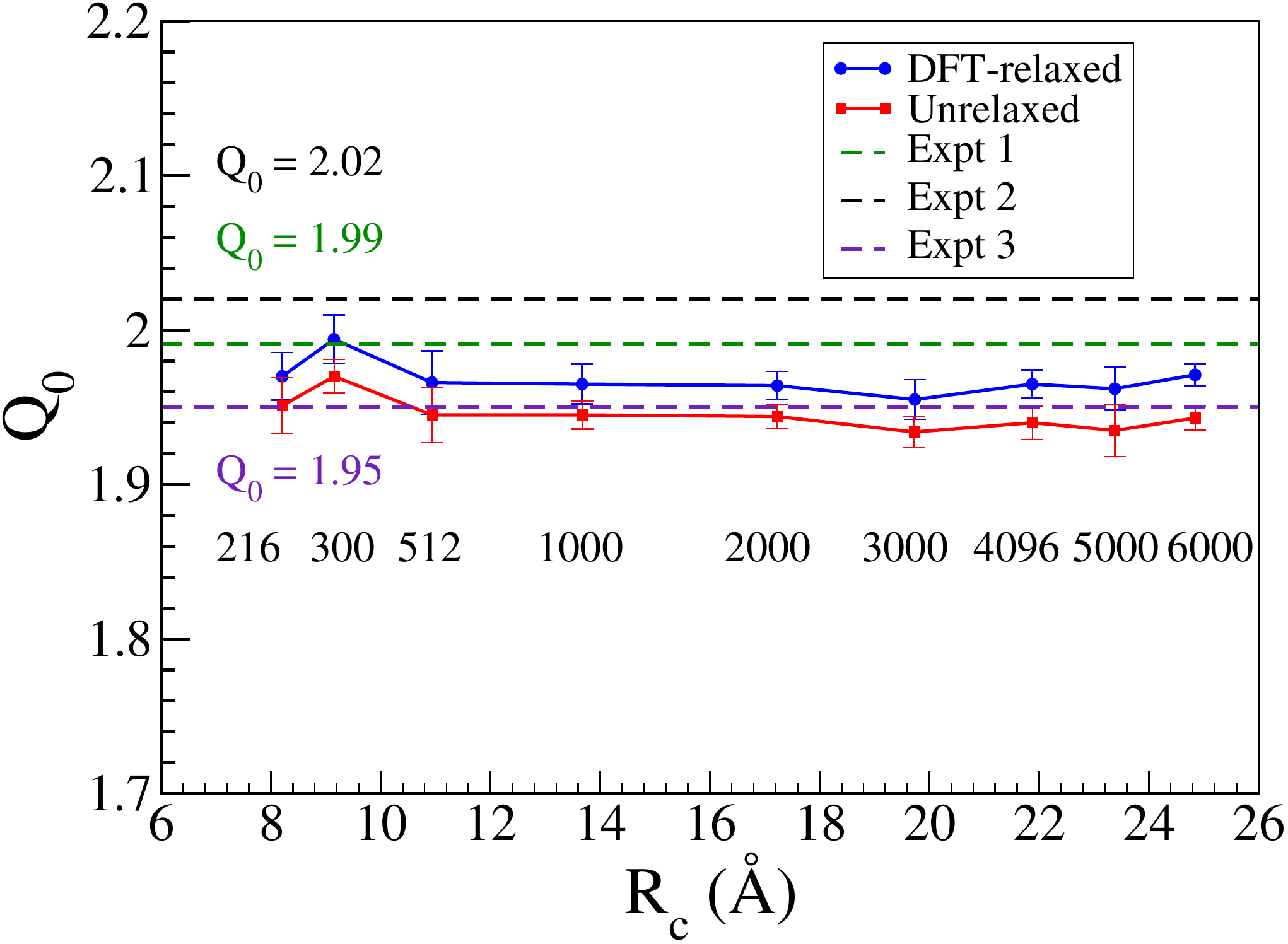}
\caption{
The dependence of the position of the FSDP, $Q_0$, with the size 
of the models before (\textcolor{red}{\tiny{$\blacksquare$}}) 
and after (\textcolor{blue}{\textbullet}) DFT relaxations. The 
horizontal lines correspond to the experimental value 
of $Q_0$ for as-implanted samples of {\asi} from Refs.\,\cite{Xie:2013} 
(green), \cite{Laaziri:1999} (black), and \cite{Fortner:1989} (indigo). 
} 
\label{fig10}
\end{figure}

The full width at half maximum, or FWHM,  of the FSDP for 
different models is plotted against $R_c$ in Fig.\,\ref{fig9}. 
A somewhat high value of the FWHM for the large DFT-relaxed 
models is a consequence of the reduction of the peak intensity. 
As the intensity of the peak reduces, the FWHM increases 
slightly due to the widening of the diffraction plot away 
from the peak.  An inspection of Figs.~\ref{fig8} 
and \ref{fig9} appears to suggest that the values of FWHM 
and $S(Q_0)$ are somewhat correlated with each other.  
In particular, amorphous-silicon models exhibiting smaller 
values of $S(Q_0)$ (in Fig.~\ref{fig8}) tend to produce somewhat 
larger values of FWHM in Fig~\ref{fig9}, irrespective of 
the size of the models and DFT relaxation. 
This is apparent in Fig.~\ref{extra}, where FWHM and $S(Q_0)$ 
values for all configurations and sizes are presented in the 
form of a scattered plot. A simple analysis of FWHM and $S(Q)$ data by 
computing the Pearson correlation coefficient, 
$r_{XY}$, confirms the suggestion that FWHM and $S(Q_0)$ values 
are indeed linearly correlated with each other and 
have a correlation coefficient of $r_{XY} = -0.9$.  The 
linear least-square (LS) fit of the data are also 
shown in Fig.~\ref{extra} by a solid (red) line.  
The great majority of the FWHM and 
$S(Q)$ values in Fig.~\ref{extra} can be seen to 
have clustered along the straight line within a 
rectangular region bounded by the experimental 
values of FWHM and $S(Q_0)$, from 0.54 to 
0.77 {\AA}$^{-1}$ and 1.37 to 1.55, 
respectively.  
Likewise, the dependence of the position 
of the FSDP with $R_c$ for the unrelaxed and DFT-relaxed models 
is illustrated in Fig.\ref{fig10}. For the unrelaxed models, 
$Q_0$ is observed to converge near 1.96 {\AA}$^{-1}$, whereas 
the corresponding value for the DFT-relaxed models hovers 
around 1.97 {\AA}. In both the cases, $Q_0$ is within 
the range of the experimental values, from 1.95 {\AA}$^{-1}$ 
to 2.02 {\AA}$^{-1}$, shown in Fig.~\ref{fig10}. 

In summary, a systematic study of {\asi} models, 
consisting of 216 to 6000 atoms, firmly establishes 
that the structure of the FSDP in {\asi} 
is mostly determined by radial pair correlations 
up to a distance of 15 {\AA}, as 
far as the size of the largest models employed 
in this study is concerned. Further, the 
major contribution to the FSDP arises from 
the second and fourth radial shells, along 
with small residual contributions from the 
distant radial shells at a distance of up to 
15 {\AA}.

\subsection{The FSDP and the radial shell structures of {\asi}}

Earlier, in sections IIIB and IIIC, we have demonstrated that the position 
of the FSDP, $Q_0$, is primarily determined by $F_2(Q)$ and, 
to a lesser extent, $F_4(Q)$. This leads to a possibility 
of the existence of a simple functional relationship 
between $Q_0$ and a suitable length scale in the real space 
involving the radial atomic correlations in the network.  
In this section, we will show that an {\em approximate} 
relationship between $Q_0$ and the average radial 
distance, $\langle R_2 \rangle$, of the atoms in the 
second (radial) shell does exist. Below, we first 
provide a rationale behind the origin of this 
relationship, which is subsequently corroborated by 
results from direct numerical calculations. 

\begin{figure}[t!]
\centering
\includegraphics[width=.4\textwidth]{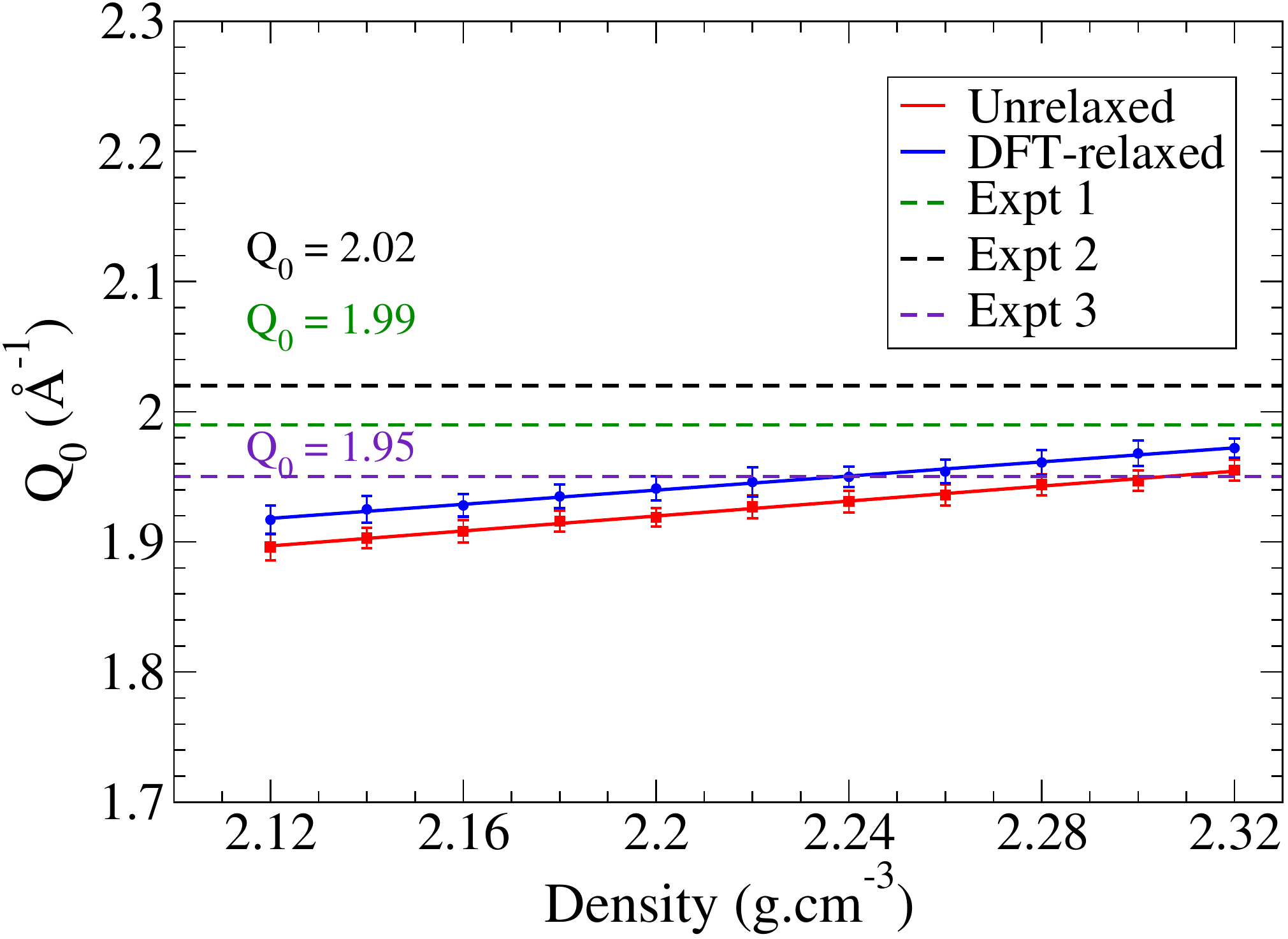}
\caption{
The variation of the peak position ($Q_0$) for 3000-atom 
models of {\asi} with its mass density before 
(\textcolor{red}{\tiny{$\blacksquare$}}) and after 
(\textcolor{blue}{\textbullet}) DFT relaxations. 
The value of $Q_0$ has been observed to vary linearly 
with the density of the model. The experimental values 
of $Q_0$ (horizontal dashed lines) correspond to 
as-implanted samples of density 2.28 g.cm$^{-3}$ from 
Refs.~\cite{Laaziri:1999} (black), \cite{Xie:2013} (green), 
and \cite{Fortner:1989} (indigo). 
} 
\label{fig11}
\end{figure}

The first hint that an approximate relationship may exist 
follows from the behavior of $Q_0$ with the (mass) density, 
$\rho$, of the models.  In Fig.~\ref{fig11}, we have 
plotted the variation of $Q_0$ against $\rho$ for {\asi}. 
For this purpose, the density of a set of 3000-atom models 
is varied, within the range from 2.12 g.cm$^{-3}$ 
to 2.32 g.cm$^{-3}$, 
by scaling the length of the cubic simulation cell and the 
position of the atoms therein. This involves a tacit assumption 
that for a small variation of the density, by about $\pm$5\%, 
the atomistic structure of the network would remain 
unchanged and that a simple scaling approach 
should suffice to generate low/high-density models. Given 
that the WWW models of {\asi} do not include any 
extended defects and voids in the network, the scaling 
assumption is reasonably correct and suitable to 
produce models with a small variation of the density. 
Figure \ref{fig11} presents the results from our 
calculations, which show a linear relationship between 
$Q_0$ and the density, $\rho$, of the models. This 
linear variation of $Q_0$ with $\rho$ is not particularly 
unique to {\asi}; a similar 
behavior has been observed experimentally by 
Inamura et al.~\cite{Inamura:1997,Inamura:2007} 
for densified silica. 

The results from Fig.~\ref{fig11} and the experimental data from 
Refs.~\cite{Inamura:1997} and \cite{Inamura:2007} suggest 
that $Q_0$ can vary approximately 
linearly with the average density, $\rho$, of 
the models/samples. Since $\rho$ is inversely 
proportional to the cubic power of the simulation 
cell size ($L$) for a given number of atoms, 
$Q_0$ also varies as $1/L^3$ when the density 
is varied by rescaling the volume. Thus, for homogeneous
and isotropic models with no significant variation 
of the local density, which the WWW models satisfy 
in the absence of extended defects and voids, it 
is reasonable to assume that $Q_0 \propto 1/r_{ij}^3$, 
where $r_{ij}(\rho)$ is the distance between any two 
atoms in the network, at sites $i$ and $j$, 
of average density $\rho$.  In view 
of our earlier observation that the position of 
the FSDP is largely determined by $F_2(Q)$(see 
Fig.~\ref{fig6}), one may posit that $r_{ij}$ 
values between $R_2$ and $R_3$ in $G(r)$ mostly 
affect the peak position at $Q_0$. These considerations 
lead to the suggestion that by substituting 
$r^3_{ij}$ by its average value of 
$\langle r^3_{ij} \rangle = \lena$ 
for the atoms in the second radial shell, 
$Q_0\lena$ should remain constant, 
on average, upon density variations via volume 
rescaling. 
Likewise, one can invoke the same reasoning and 
may expect $Q_0\lenb$ should be also constant 
but only approximately, due to the limited role 
and contribution of the atoms in the fourth 
radial shell in determining the position of 
$Q_0$.

\begin{figure}[t!]
\centering
\includegraphics[width=.4\textwidth]{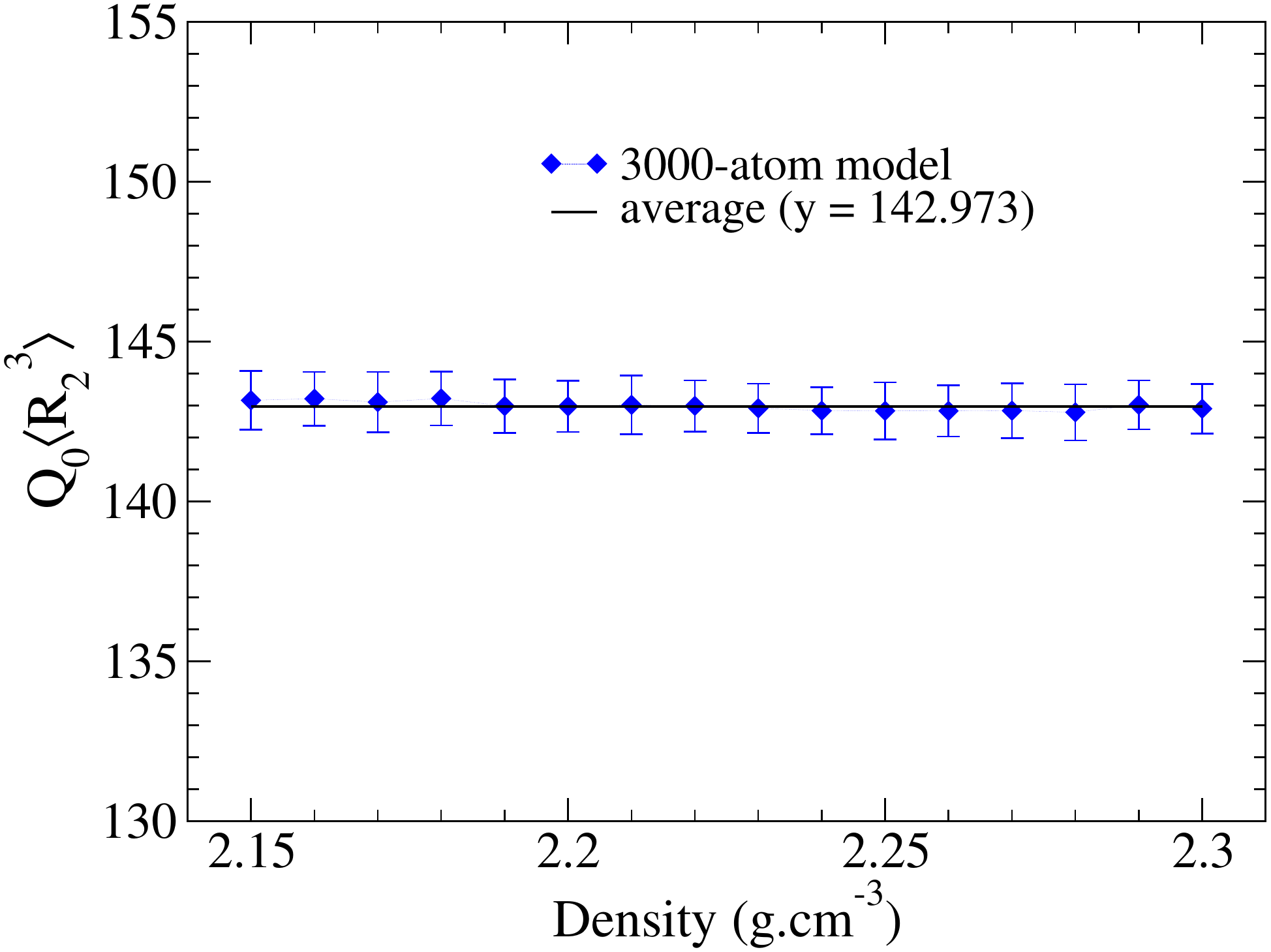}
\caption{
The relation between $Q_0$ and $\lena$ of the 
atoms in the second radial shell. A constant 
value of $Q_0{\lena}$ with respect to the 
density of {\asi} models indicates that $Q_0$ 
is approximately proportional to the inverse 
of $\lena$. The horizontal black line indicates 
the average value of $Q_0{\lena}$ within the 
density range shown in the plot.  
}
\label{fig12}
\end{figure}

\begin{figure}[t!]
\centering
\includegraphics[width=.4\textwidth]{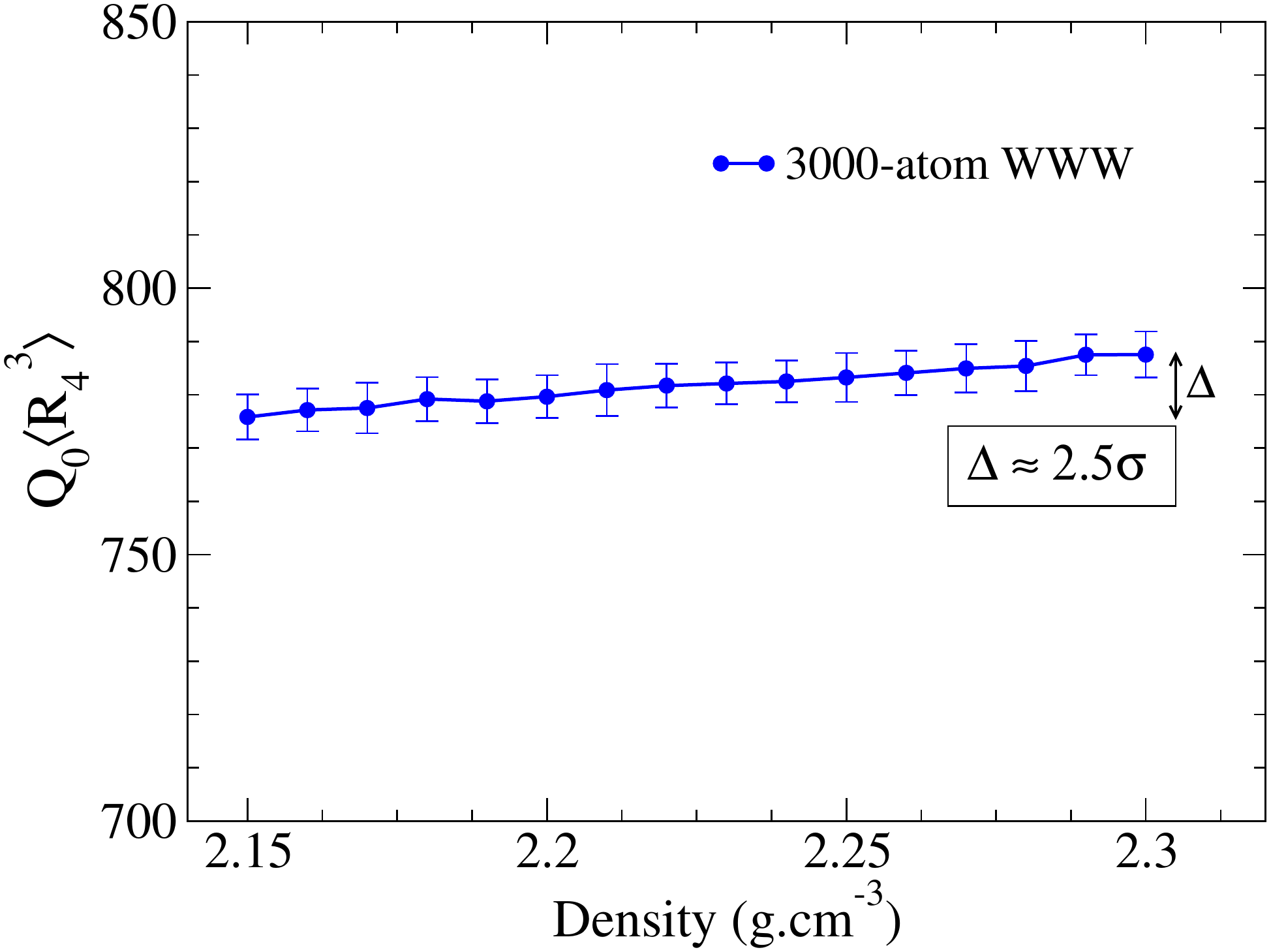}
\caption{
The dependence of $Q_0$ and $\lenb$ for the atoms in 
the fourth radial shell of {\asi} in the density 
range from 2.15 to 2.3 g.cm$^{-3}$. The large 
deviation of $Q_0\lenb$ values, indicated by 
$\Delta \approx 2.5\,\sigma$, from a constant 
value suggests that no simple relationship between 
$Q_0$ and $\lenb$ exists. 
}
\label{fig13}
\end{figure}

The efficacy of our argument can be verified by 
results from direct numerical calculations.  A plot of 
$Q_0\lena$ (and $Q_0\lenb$) versus the average 
density $\rho$ in Fig.~\ref{fig12} 
(and Fig.~\ref{fig13}) indeed confirms our prediction. 
It may be noted that the observed (absolute) deviation, 
$\Delta$, of $Q_0{\lena}$ values in 
the density range 2.15--2.3 g.cm$^{-3}$ in 
Fig.~\ref{fig12} is of the order of $\pm 0.46\,\sigma$, 
where $\sigma$ is the (largest) standard deviation obtained 
by averaging results from 10 independent models for 
each density.  By contrast, the corresponding deviation 
for $Q_0{\lenb}$ in Fig.~\ref{fig13} 
is found to be more than two standard deviation, as 
indicated in the plot. The large deviation of 
$Q_0{\lenb}$ values is not unexpected 
in view of the small contribution of $F_4(Q)$ (to the 
FSDP) that originates from the fourth radial shell.  
Thus, the results from Fig.~\ref{fig12} lead to the 
conclusion that $Q_0$ is approximately proportional 
to the inverse of the average cubic power of the 
radial distance, $\lena$, of the atoms in the second 
radial shell in {\asi}.  
It goes without saying that 
the use of ${\langle R_2 \rangle}^3$, instead of 
${\lena}$, does not change the conclusion of our 
work, as the difference between these two values 
is found to be about 1.92--2.1 {\AA}$^3$, for 
the mass density in the range of 2.15 to 2.3 
g.cm$^{-3}$, which simply shifts the plot 
(in Fig.~\ref{fig12}) vertically downward by a 
constant amount. 

\begin{figure}[ht!]
\centering
\includegraphics[width=.4\textwidth]{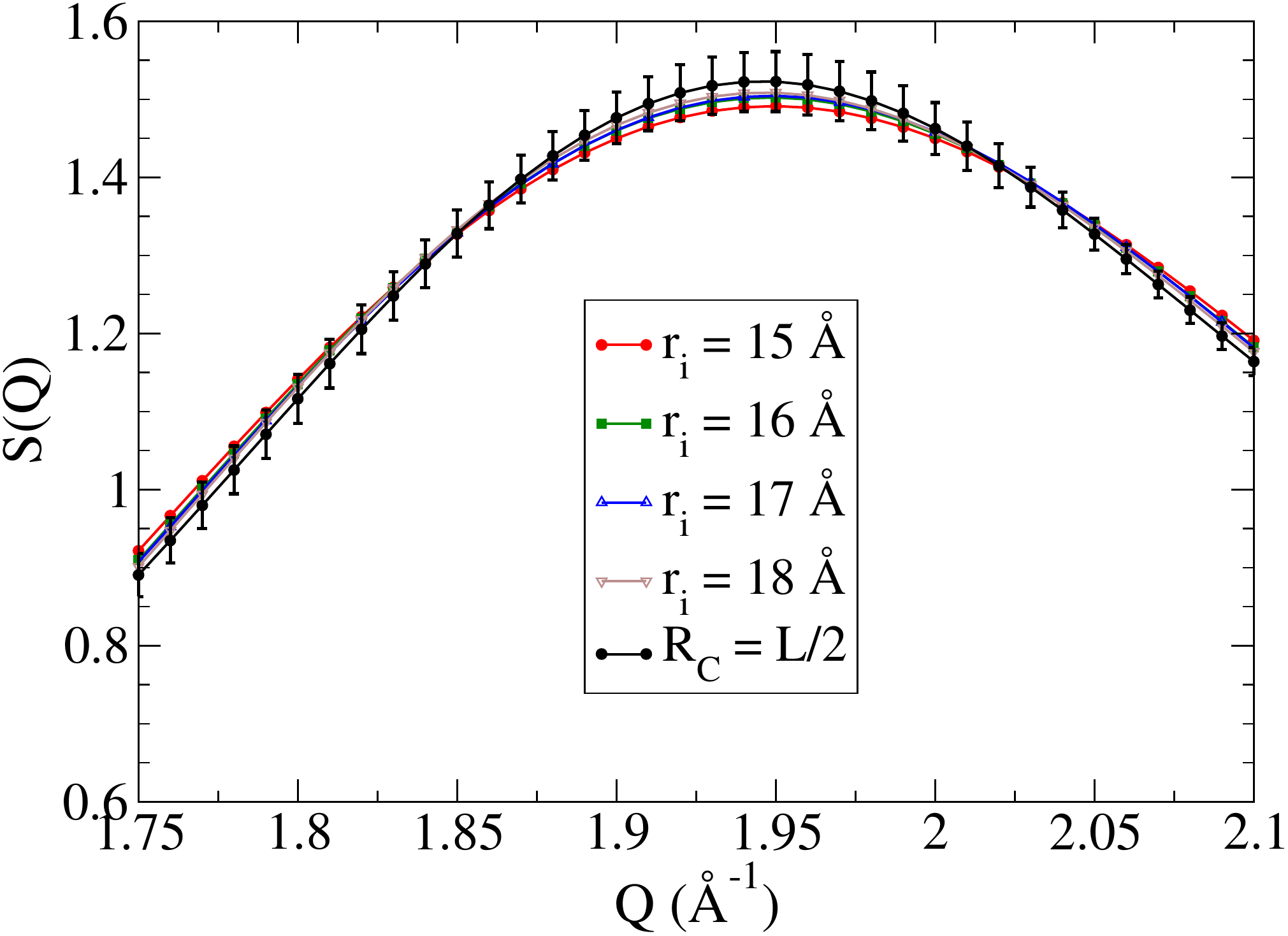}
\caption{
The structure factor, $S(Q)$, in the vicinity of the 
FSDP from 6000-atom {\it a}-Si models. The change in 
$S(Q)$ due to varying radial cutoff distances, $r_i$, 
is found to be less than one standard deviation 
($\Delta S(Q_0) \approx 0.85\,\sigma$ for $r_i$ = 15 {\AA}, 
and 0.46$\,\sigma$ for $r_i$ = 18 {\AA}). The 
standard deviation, $\sigma$, is obtained from using 
the maximal radial cutoff $L/2$, which is given by 
24.85 {\AA}.  The results correspond to the average 
values of $S(Q)$ obtained from 10 configurations.  
} 
\label{fig14}
\end{figure}

\begin{figure}[t!]
\centering
\includegraphics[width=.4\textwidth]{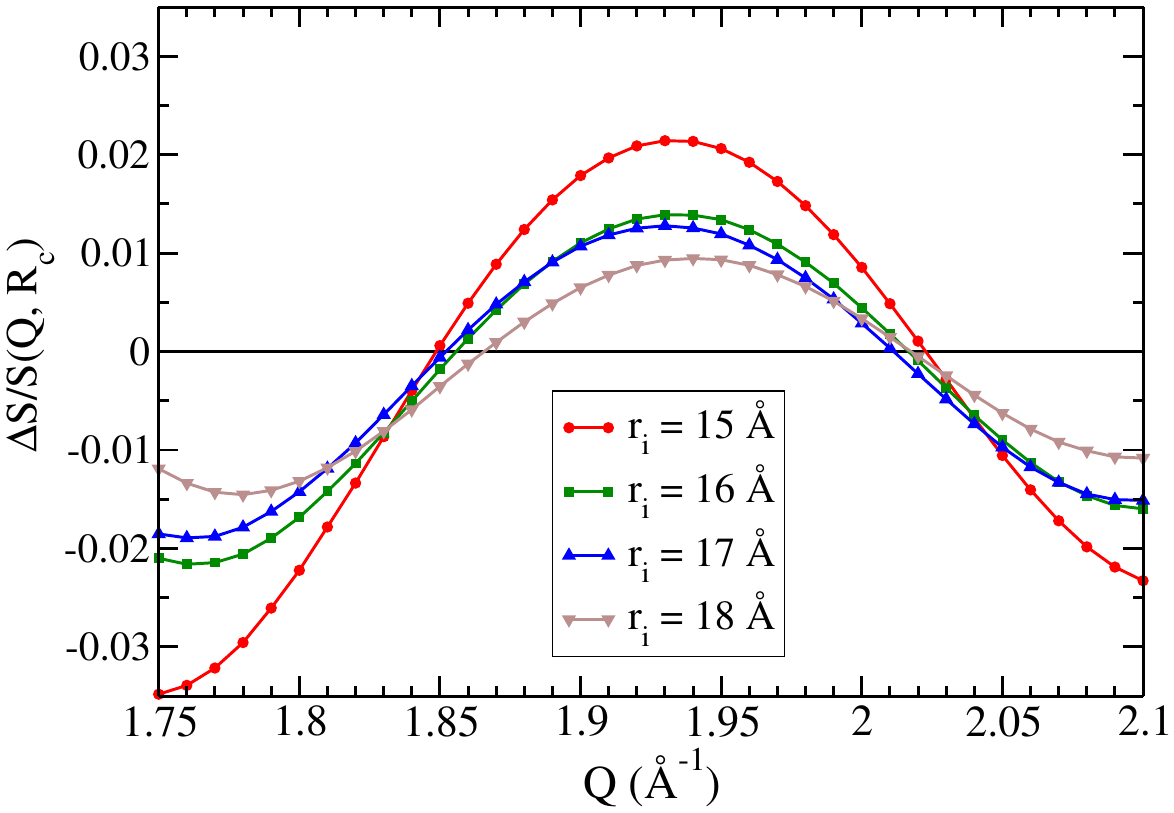}
\caption{
The fractional error associated with the calculation 
of the structure factor in the vicinity of the FSDP 
with a varying radial cutoff distance, $r_i$, from 
15 {\AA} to 18 {\AA}.  $\Delta S(Q)$ is the absolute 
error and $R_c$ (=24.85 \AA) is the half-length of 
the cubic simulation cell for 6000-atom models. 
The error due to the truncation of the radial distance 
at $r_i$ can be seen to be around 1--2\%, which is well 
within one standard deviation of $S(Q_0)$ 
(see Fig.~\ref{fig14}). 
} 
\label{fig15}
\end{figure}

We end this section by making a comment on the possible 
role of distant radial atomic correlations, or 
extended-range oscillations (ERO), in $G(r)$ on the 
FSDP, based on our preliminary results from 6000-atom 
models. Although the presence of ERO in ultra-large models 
of {\asi} beyond 15 {\AA} is an undisputed 
fact~\cite{Uhlherr:1994}, a direct determination of 
the effect of the ERO on the FSDP in {\asi} is highly 
nontrivial due to the presence of intrinsic noises 
in $G(r)$ at large radial distances.  Numerical 
calculations using 6000-atom models of {\asi} 
indicate that only a minute fraction of the total intensity 
of the FSDP results from the radial region beyond 
15 {\AA}. These calculations do not include 
any possible artifacts that may arise from the 
noises in $G(r)$ at large distances.
The observed deviation in the peak intensity, 
due to the truncation of the radial distance at 15 {\AA} and 
at higher values, is found to be about 1--2\%, 
which is less than one standard deviation ($\sigma$) 
associated with $S(Q_0)$, obtained from using the 
maximal radial cutoff distance $R_c(=L/2)$, as far 
as the results from 6000-atom models are concerned. 
Figure \ref{fig14} shows the variation of the intensity 
near the FSDP for five different cutoff values from 
15 {\AA} to 18 {\AA} and 24.85 {\AA}. It is apparent 
that the changes in $S(Q)$ near $Q_0$ are very small as 
the radial cutoff value increases from 15 {\AA} to 
18 {\AA}. These small changes in the intensity values 
are readily reflected in Fig.~\ref{fig15}, where the 
fractional errors, with respect to $S(Q,R_c)$,  
associated with the calculation of $S(Q,r_i)$ are 
plotted against $Q$ for $r_i$ = 15 {\AA} to 18 {\AA}. 
Thus, as far as the present study and the maximum 
size of the models are concerned, the ERO do not 
appear to contribute much to the FSDP. However, an accurate 
study of the ERO in {\asi} would require high-quality 
ultra-large models, consisting of several tens of thousands 
of atoms, and a suitable prescription to handle 
noises in $G(r)$ at large distances. These and some 
related issues concerning the origin of the ERO in 
{\asi} and their possible role in $S(Q)$ will be 
addressed in a future communication. 

\section{Conclusions}
\label{S:4}

In this paper, we have studied the origin and structure 
of the FSDP of {\asi} with an emphasis on the position, intensity, and 
width of the diffraction peak. The study leads to the 
following results: 
1) By partitioning the contribution of the reduced PCF 
to the FSDP, which originates from the Fourier transform 
of radial atomic correlations in the real space, a 
quantitative measure of the contribution to the FSDP 
from different radial shells is obtained. The results 
show that the position of the FSDP in {\asi} is principally 
determined by atomic pair correlations in the second, 
fourth, and sixth radial shells, in the descending order 
of importance, supplemented by small residual contributions 
from beyond the sixth radial shell; 
2) A convergence study of the position, intensity, 
and width of the FSDP, using a set of models of size 
from 216 to 6000 atoms, suggests that the minimum 
size of the models must be at least 1000 atoms or more 
in order for the results to be free from finite-size 
effects.  This approximately translates into a radial 
length of 14 {\AA}, which is consistent with the 
results obtained from the radial-shell analysis 
of the reduced PCF; 
3) A theoretical basis for the results obtained from 
numerical calculations is presented by examining 
the relationship between the peaks in the structure 
factor and the reduced PCF.  
Contrary to the common assumption that the peaks in 
the structure factor and the reduced PCF are not 
directly related to each other, we have shown 
explicitly that the knowledge of the reduced PCF 
alone is sufficient not only to determine the approximate 
position of the FSDP and the principal peak but also 
the relevant radial regions that are 
primarily responsible for the emergence of these 
peaks in the structure factor, and vice versa; 
4) The study leads to an approximate relation 
between the position of the FSDP and the average 
radial distance of the atoms in the second 
radial shell of {\asi} networks. For homogeneous 
and isotropic models of {\asi} with no significant 
variation of the local density, it has been shown 
that the position of the FSDP is inversely proportional 
to the cubic power of the average radial distance 
of the atoms in the second radial shell.
The result is justified by providing a phenomenological 
explanation -- based on experimental and computational 
studies of the variation of the FSDP with the average 
density of {\asi} samples and models --  which is 
subsequently confirmed by direct numerical calculations 
for a range of density from 2.15 to 2.3 g.cm$^{-3}$. 
\vspace*{0.5 cm} 

\section*{Acknowledgements}
The work was partially supported by the U.S.~National 
Science Foundation (NSF) under Grant No.~DMR 1833035. 
One of us (P.B.) thanks Profs.~Gerard Barkema (Utrecht, 
The Netherlands) and Normand Mousseau (Montreal, 
Canada) for providing their modified WWW code.  
\bibliographystyle{apsrev4-1}

\begin{thebibliography}{45}
\bibitem [{BMn()}]{BMnote}%
  \BibitemOpen
  \href@noop {} {}\bibinfo {note} {An alternative approach to generate a
  4-fold-coordinated disordered network is to employ the reverse Monte Carlo
  (RMC) approach using a suitable cost function. The resulting tetravalent
  network can be used as a starting configuration for the WWW bond switches
  between atoms.}\BibitemShut {Stop}%
\bibitem [{sin()}]{sinc}%
  \BibitemOpen
  \href@noop {} {}\bibinfo {note} {The positions of the maxima of $\sin(x)/x$
  asymptotically converge to $(4m+1)\pi/2$, where $m$ is a positive integer.
  The first four positions of the maxima of $\sin(x)/x$, for $x \ge 0$, are
  given by $x$=0, 7.725, 14.065, and 20.37. The corresponding values for
  $\sin(x)$ are at $x$=1.570 ($\pi/2$), 7.853 ($5\pi/2$), 14.137 ($9\pi/2$),
  and 20.420 ($13\pi/2$). This also explains why $m$=0 does not provide a
  physical solution (as it corresponds to $Qr$ = 0).}\BibitemShut {Stop}%
\end{thebibliography}
%

\end{document}